\documentclass[10pt,journal,compsoc]{IEEEtran}
\usepackage{amsmath,amssymb,amsfonts}
\usepackage{bm}
\usepackage[nocompress]{cite}
\usepackage{graphicx}
\usepackage{amsmath}
\usepackage{algorithm}
\usepackage{algpseudocode} 
\usepackage{array} 
\usepackage{enumitem}
\usepackage{booktabs} % For professional table formatting

\usepackage{siunitx}  % For scientific notation and number formatting

\usepackage[caption=false,font=footnotesize,labelfont=sf,textfont=sf]{subfig}
\usepackage{textcomp}
\usepackage{stfloats}
\usepackage{url}
\usepackage{verbatim}

\usepackage[dvipsnames]{xcolor}
\usepackage{diagbox}
\usepackage{multirow}
\begin{document}
\title{Look Before Switch: Sensing-Assisted Handover in 5G NR V2I Networks}

\author{Yunxin~Li,~\IEEEmembership{Graduate~Student~Member,~IEEE,}
        Fan~Liu,~\IEEEmembership{Senior~Member,~IEEE,}
        Haoqiu~Xiong,~\IEEEmembership{Graduate~Student~Member,~IEEE,}
        Zhenkun~Wang,~\IEEEmembership{Senior~Member,~IEEE,}
        Narengerile,~\IEEEmembership{Member,~IEEE,}
        and~Christos~Masouros,~\IEEEmembership{Fellow,~IEEE}% <-this % stops a space
\IEEEcompsocitemizethanks{
\IEEEcompsocthanksitem %This work was supported in part by the National Natural Science Foundation of China under Grant 62101234, 62331023 and Grant 62301264, in part by the Guangdong Province "Pearl River" Young Talent Support Program under Grant 2021QN02X128, in part by the Shenzhen Science and Technology Program under Grant RCBS20210609103227018 and Grant 20220815100308002, in part by the Natural Science Foundation of Jiangsu Province under Grant BK20230416, and in part by the major key project of Peng Cheng Laboratory under grant PCL2023AS1-2. 
A part of this article was presented at the IEEE International Conference on Communications (ICC), Canada, June 2025.% <-this % stops a space
\IEEEcompsocthanksitem Yunxin Li is with the School of Automation and Intelligent Manufacturing (AIM), Southern University of Science and Technology, Shenzhen 518055, China, and also with the Department of Electrical Engineering (ESAT), KU Leuven, 3001 Leuven, Belgium.\protect\\
E-mail: liyx2022@mail.sustech.edu.cn.
\IEEEcompsocthanksitem Fan Liu is with the School of Automation and Intelligent Manufacturing, Southern University of Science and Technology, Shenzhen 518055, China.\protect\\
E-mail: f.liu@ieee.org.
\IEEEcompsocthanksitem Haoqiu Xiong is with the Department of Electrical Engineering, KU Leuven, 3001 Leuven, Belgium. \protect\\
E-mail: haoqiu.xiong@kuleuven.be.
\IEEEcompsocthanksitem Zhenkun Wang is with the School of Automation and Intelligent Manufacturing, Southern University of Science and Technology, Shenzhen 518055, China. \protect\\
E-mail: wangzhenkun90@gmail.com.
\IEEEcompsocthanksitem Narengerile is with the Wireless Technology Lab,
Huawei Technologies Company Ltd., Shenzhen 518129, China. \protect\\
E-mail: narengerile@huawei.com.
\IEEEcompsocthanksitem Christos Masouros is with the Department of Electronic and Electrical Engineering, University College London, London WC1E 7JE, U.K. \protect\\
E-mail: c.masouros@ucl.ac.uk.
}% <-this % stops a space
\thanks{(Corresponding author: Fan Liu.)}
}

\IEEEtitleabstractindextext{%
\begin{abstract}

Integrated Sensing and Communication (ISAC) has emerged as a promising solution in addressing the challenges of high-mobility scenarios in 5G NR Vehicle-to-Infrastructure (V2I) communications. This paper proposes a novel sensing-assisted handover framework that leverages ISAC capabilities to enable precise beamforming and proactive handover decisions. Two sensing-enabled handover triggering algorithms are developed: a distance-based scheme that utilizes estimated spatial positioning, and a probability-based approach that predicts vehicle maneuvers using interacting multiple model extended Kalman filter (IMM-EKF) tracking. The proposed methods eliminate the need for uplink feedback and beam sweeping, thus significantly reducing signaling overhead and handover interruption time. A sensing-assisted NR frame structure and corresponding protocol design are also introduced to support rapid synchronization and access under vehicular mobility. Extensive link-level simulations using real-world map data demonstrate that the proposed framework reduces the average handover interruption time by over 50\%, achieves lower handover rates, and enhances overall communication performance.
\end{abstract}

}

\maketitle
\IEEEdisplaynontitleabstractindextext
\IEEEpeerreviewmaketitle
\ifCLASSOPTIONcompsoc
\IEEEraisesectionheading{\section{Introduction}\label{sec:introduction}}
\else
\section{Introduction}
\label{sec:introduction}
\fi

\IEEEPARstart{T}{he} rapid evolution of intelligent transportation systems (ITS) is fundamentally reshaping modern mobility, driven by advancements in wireless communication technologies such as 5G-Advanced (5G-A) and the anticipated 6G. At the heart of this transformation lies Vehicle-to-Everything (V2X) communication, a paradigm that enables vehicles to connect seamlessly with their surroundings, encompassing infrastructure (V2I), pedestrians (V2P), other vehicles (V2V), and network entities (V2N). This interconnected ecosystem promises to enhance road safety, optimize traffic efficiency, and pave the way for fully autonomous driving~\cite{chen2017vehicle,boban2018connected}. With the advent of 5G New Radio (NR), V2X networks leverage ultra-low latency, high reliability, and massive device connectivity to support next-generation vehicular applications such as adaptive cruise control and vehicle platooning~\cite{clancy2024wireless}.

% Detailing V2X technologies: DSRC and C-V2X
Two primary technologies underpin V2X communications: Dedicated Short-Range Communications (DSRC) and Cellular V2X (C-V2X), each with distinct characteristics and challenges. DSRC, rooted in the IEEE 802.11p standard and operating in the 5.9 GHz band, has been a cornerstone of early V2X deployments, offering low-latency, ad-hoc communication suited for safety-critical applications like collision warnings~\cite{kenney2011dedicated,yin2004performance}. However, DSRC faces significant limitations: its modest data rates (ranging from 3 to 27 Mbps) and restricted coverage (typically 300–1000 meters) struggle to support the bandwidth demands of modern V2X services, especially in dense urban environments or highway scenarios with high vehicle speeds~\cite{abboud2016interworking,moradi2023dsrc}. Moreover, its small cell size exacerbates handover (HO) frequency, leading to increased latency and potential connectivity disruptions as vehicles transition between access points. In contrast, C-V2X, evolving from LTE-V2X to 5G NR-based V2X, harnesses cellular infrastructure to deliver higher data rates, broader coverage, and improved scalability~\cite{molina2017lte,chen2016lte}. C-V2X supports both sidelink and network-based communications, making it versatile for diverse use cases~\cite{chen2020vision}.

% Focusing on handover challenges in V2X networks
Unlike traditional cellular networks, where users move at pedestrian speeds, V2X networks involve vehicles traveling at velocities up to 120 km/h or more, crossing cell boundaries in mere seconds. This high mobility makes handovers between gNBs critical for ensuring uninterrupted connectivity in 5G NR V2X systems~\cite{gyawali2020challenges,clancy2023investigating}. Prior work on handover in NR and V2X systems has primarily focused on communication-only approaches, relying on metrics like Reference Signal Received Power (RSRP) or Reference Signal Received Quality (RSRQ) to trigger handover decisions~\cite{3gpp.38.331, garcia2021tutorial}. These methods assess signal strength to select the optimal gNB, often supplemented by predictive algorithms that leverage vehicle trajectory data or historical mobility patterns to anticipate handover needs. For instance, machine learning-based predictive models have been proposed to estimate future vehicle positions and optimize handover timing, improving efficiency in stable environments~\cite{qi2020federated,koda2019handover}. However, these approaches face significant challenges in dynamic, high-mobility V2I scenarios. Rapid channel variations due to fading, millimeter-wave (mmWave) beam misalignment, and environmental blockages (e.g., buildings or other vehicles) frequently lead to handover failures or suboptimal decisions. Additionally, the reliance on communication-only metrics introduces latency in processing real-time channel changes, degrading quality of service (QoS) and failing to meet the ultra-low latency requirements of safety-critical V2X applications \cite{zaidi2020mobility}. These limitations highlight the need for more robust handover strategies that can adapt to the dynamic and unpredictable nature of V2I environments.

To address the limitations of traditional vehicular networks, Integrated Sensing and Communications (ISAC) has emerged as a revolutionary paradigm~\cite{liu2022integrated}. By co-designing sensing and communication functionalities within a unified framework, ISAC leverages shared spectrum and hardware resources, such as mmWave bands and massive MIMO (mMIMO) arrays, to achieve \textit{integration gain} through efficient spectrum reuse and \textit{coordination gain} through joint optimization of sensing and communication performance~\cite{cui2021integrating, zheng2019radar}. This dual-purpose approach aligns with the vision of IMT-2030, which identifies ISAC as a key usage scenario for 6G networks~\cite{ITU-R-M2160}. On top of that, mMIMO and mmWave further enhance ISAC by providing high-capacity, directional beams to meet the stringent throughput and latency demands of NR V2I systems. By providing environmental awareness through sensing, ISAC offers a novel approach to improve the communication service, known as \textit{sensing-assisted communications}. 

Prior ISAC research has largely concentrated on foundational signal processing and beamforming design, aiming to achieve precise beam tracking for high-mobility scenarios. Pioneering work introduced predictive beamforming and beam tracking, leveraging dual-functional radar-communication (DFRC) systems to estimate vehicle trajectories with minimal overhead, using techniques such as Bayesian message passing~\cite{yuan2020bayesian} and extended Kalman filtering~\cite{liu2020radar}, while addressing complex scattering from extended vehicle targets~\cite{du2022integrated}. The integration of deep learning has further elevated ISAC systems, with adaptive beamforming strategies learning from historical data to navigate dynamic urban environments, as demonstrated in real-world 6G experiments~\cite{mu2021integrated, liu2022learning, demirhan2022radar}. To manage dense networks, multi-beam and mMIMO approaches enable simultaneous tracking of multiple vehicles, improving spectral efficiency~\cite{xiao2024simultaneous, chen2023multiuser}, while cell-free ISAC systems enhance coverage through distributed beamforming~\cite{demirhan2024cell}. More recently, hardware-reconfigurable architectures such as fluid antennas (FAS) and reconfigurable intelligent surfaces (RIS) have shown potential to support dynamic beam steering and joint sensing-communication optimization for 6G networks~\cite{yang2025towards, yu2023active, zhou2024fluid}. These developments collectively advance sensing-assisted communications by optimizing beam tracking and link reliability for autonomous and connected vehicles.

Despite these significant advances, most ISAC studies remain focused on designing algorithms for beam alignment and trajectory prediction, with limited consideration of how sensing information can be integrated into standardized 5G NR protocols. Existing ISAC-aided mobility models typically enhance tracking accuracy but do not address protocol-level aspects such as measurement reporting, triggering, or signaling compatibility with 3GPP procedures. In contrast, traditional communication-only handover schemes depend solely on received power or quality metrics, overlooking the potential of sensory data to provide environmental awareness. To bridge this gap, this work proposes a sensing-assisted handover framework that integrates sensing information into NR-compliant handover processes.

It is important to note that, although ISAC has been widely studied from a signal processing perspective, its standardization is still at an early stage. 3GPP has recently initiated ISAC-related activities beginning with Release~18, which enhanced positioning capabilities as a preliminary step toward wireless sensing in 5G-Advanced. TR~22.837~\cite{3gpp.22.837} in Release~19 identified over thirty ISAC use cases, followed by TS~22.137~\cite{3gpp.22.137} that defined key performance indicators for sensing services, such as positioning accuracy, velocity estimation accuracy, and sensing latency. While these activities have established high-level requirements, normative procedures for sensing-based beam management, synchronization, or handover have not yet been introduced in current NR releases (up to Release~18). Therefore, by reusing NR-compliant frame and protocol structures, this work demonstrates how ISAC functionality can be progressively integrated into future 5G-A and 6G systems, offering both a technically feasible and standard-compatible reference for real-world implementation.

To address the aforementioned challenges, this paper proposes a novel sensing-assisted handover framework for 5G NR V2I networks, overcoming the limitations of existing ISAC and communication-only approaches. The proposed framework integrates sensing-assisted triggering algorithms, tailored frame structures, and protocols fully compatible with 5G NR standards, designed to minimize reference signal overhead, reduce decision-making latency, and shorten interruption time during handovers. Through extensive link-level simulations incorporating real-world urban scenarios derived from OpenStreetMap (OSM) and diverse vehicle routes, the effectiveness of the framework is rigorously validated. The main contributions of this work are summarized as follows:\looseness=-1 
\begin{itemize}
    \item \textbf{Sensing-Assisted Handover Triggering Algorithms.} To reduce the latency inherent in conventional handover triggering events, we propose two novel algorithms that leverage vehicle kinematic parameters, such as distance and maneuver probability, to initiate handovers. By continuously monitoring these parameters through sensing rather than relying solely on measuring the received power of periodic reference signals, the proposed algorithms significantly decrease decision-making latency, enhancing responsiveness in high-mobility V2I scenarios.\looseness=-1 
    \item \textbf{ISAC-Enhanced Frame Structures and Protocols for Handover.} Building on 5G NR frame structures, we develop ISAC-based protocols and frame designs that utilize sensing-assisted triggering events to optimize handover performance. By identifying and eliminating redundant pilot and reference signals with ISAC-derived environmental awareness, the proposed framework reduces interruption time during handovers while maintaining full compatibility with 5G NR standards and preserving core system functionalities.\looseness=-1 
\end{itemize}

\begin{figure*}[htbp]
\centering
\includegraphics[width=18cm]{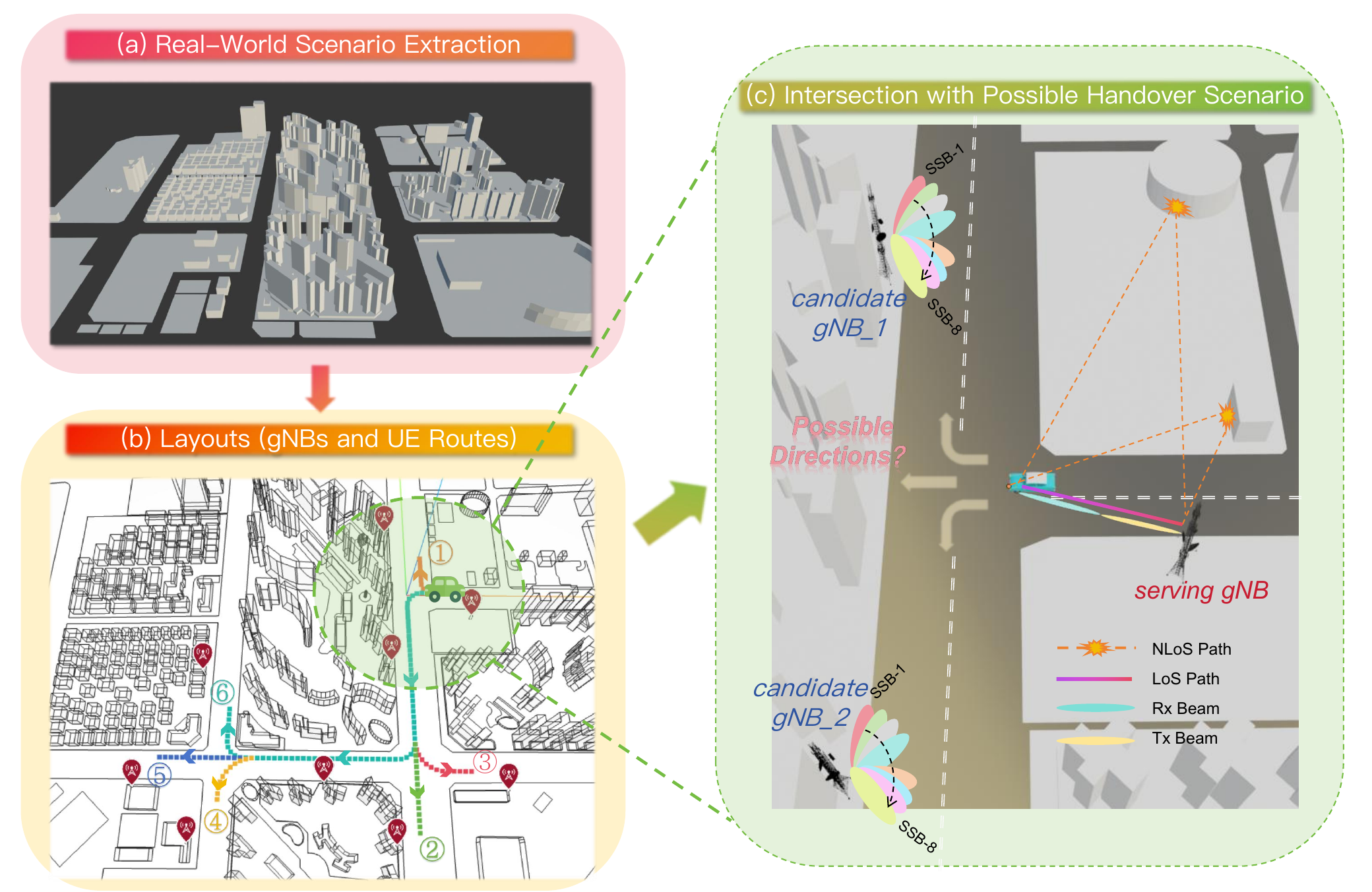}
\caption{Simulation scenario.}
\label{scenario}
\end{figure*}

This work introduces an innovative approach to handover management in 5G NR V2I networks by leveraging ISAC capabilities. The proposed sensing-assisted handover framework reduces the average handover interruption time by over 50\%, from approximately 43 ms in communication-only schemes to around 20 ms, enabling seamless connectivity for latency-sensitive vehicular applications. Additionally, the proposed framework achieves more stable handover rates by utilizing count-based triggering mechanisms. Furthermore, it delivers consistently higher throughput and lower outage probabilities, ensuring robust and reliable communication. The remainder of this paper is organized as follows, Section.~\ref{sigmodel} introduces the system models in sensing-assisted NR-V2I networks, Section.~\ref{classicHO} describes frame structures and protocols in NR classic handover, Section.~\ref{sensHO} proposes sensing-assisted tracking and triggering algorithms and its corresponding frameworks, Section.~\ref{linkSim} provides numerical results from link-level simulations in different scenarios, and finally Section.~\ref{Conclu} concludes the paper.\looseness=-1

\textbf{Notations:} Without particular specification, matrices, vectors and scalars are denoted by bold uppercase letters (i.e., $\mathbf{A}$), bold lowercase letters (i.e., $\mathbf{a}$) and normal font (i.e., $N$), respectively. $[\mathbf{a}]_{i}$ represents the $i$th element of the vector $\mathbf{a}$. $(\cdot)^*$, $(\cdot)^T$, $(\cdot)^H$ and $(\cdot)^{-1}$ represent the conjugate, transpose, Hermitian and inverse operators, respectively. $\otimes$ denotes the Kronecker product. \looseness=-1

\section{Signal Model for NR-V2I Networks}\label{sigmodel}

We consider an NR-V2I network where a single vehicle, acting as the user equipment (UE), is already operating in the connected mode with its serving gNodeB (s-gNB) via one line-of-sight (LoS) path and $K-1$ non-line-of-sight (NLoS) channels due to the scatterers, as illustrated in Fig. \ref{scenario} (c). These signals are assumed to be ISAC signals, enabling simultaneous data transmission and sensing without requiring time or frequency division multiplexing. Both the gNB and the UE are equipped with uniform planar arrays (UPAs), characterized by $N_{\text{T}}^{\text{gNB}}$, $N_{\text{R}}^{\text{gNB}}$, $M_{\text{T}}^{\text{UE}}$, and $M_{\text{R}}^{\text{UE}}$, representing the number of transmit and receive antennas for the gNB and the UE, respectively. The s-gNB operates in full-duplex mode, utilizing NR OFDM signals for data transmission.

The downlink communication signal received by the UE from the s-gNB at the $n$th time slot is given by
\begin{equation}
\begin{aligned}\label{1}
c_n(t)=  \sqrt{P N_{\text{T}}^{\text{gNB}} M_{\text{R}}^{\text{UE}}}\sum_{k=1}^K  {\alpha}_{k, n} \mathbf{v}_n^T \mathbf{u}(\bm{\theta}_{k, n}) \mathbf{a}^T(\bm{\theta}_{k, n}) \mathbf{f}_{n} {s}_{n}(t) \\
+z_c(t),
\end{aligned}
\end{equation}
where $P$ is the transmit power, ${s}_{n}(t)$ denotes the transmitted NR OFDM signal, and $\alpha_{k,n}$ is the path-loss coefficient for the $k$-th path. The vector $\bm{\theta}_{k, n}=[\theta_{k, n}, \phi_{k, n}]^T$ comprises both the azimuth angle $\theta_{k, n}$ and the elevation angle $\phi_{k, n}$ of the $k$-th scatterer relative to the s-gNB. The transmit steering vector of the s-gNB and the receive steering vectors of the UE are $\mathbf{a}(\bm{\theta}_{k, n})$ and $\mathbf{u}(\bm{\theta}_{k, n})$, while $\mathbf{f}_n$ and $\mathbf{v}_n$ are the transmit and receive beamforming vectors. The term $z_c(t)$ represents additive white Gaussian noise (AWGN) with zero mean. For the communication model, we assume negligible relative delays and Doppler shifts within the time slot, as the OFDM system employs cyclic prefixes and frequency-domain equalization to mitigate these effects, ensuring robust data decoding at the UE side.

For the sensing-assisted V2I network, the s-gNB uses reflected ISAC signals for monostatic sensing to extract the kinematic parameters of the vehicle. The reflected ISAC OFDM signals received at the s-gNB from the UE via the LoS path and $K-1$ scatterers in the NLoS paths can be formulated as
\begin{equation}
\begin{aligned}\label{2}
\mathbf{r}_n(t) = \sqrt{P N_{\text{T}}^{\text{gNB}} N_{\text{R}}^{\text{gNB}}}\sum_{k=1}^K  \Tilde{\alpha}_{k, n}  e^{j 2 \pi \mu_{k, n} t} \mathbf{b}(\bm{\theta}_{k, n}) \mathbf{a}^T(\bm{\theta}_{k, n}) \\
\cdot \mathbf{f}_{n} {s}_n(t-\tau_{k,n}) + \mathbf{z}_r(t),
\end{aligned}
\end{equation}
where $\mathbf{b}(\bm{\theta}_{k, n})$ corresponds to the s-gNB's receive steering vector. The Doppler frequency and the time delay of the $k$th scatterer are represented by $\mu_{k, n}=2v_{k,n}f_{c}c^{-1}$ and $\tau_{k,n}=2d_{k, n}c^{-1}$, respectively, which are related to the radial velocity $v_{k,n}$, relative distance $d_{k,n}$ and carrier frequency $f_{c}$. The path-loss coefficient $\tilde{\alpha}_{k, n}$ encompasses the radar cross section (RCS) of the target and other scatterers, along with distance-dependent propagation losses. The term $\mathbf{z}_r(t)$ denotes the complex additive white Gaussian noise with zero mean. The assumption of $K$ paths for both communication and sensing is based on the shared physical propagation environment, where the LoS and NLoS paths correspond to the same dominant scatterers. 

For a UPA, the steering vector $\mathbf{a}(\bm{\theta})$ (and similarly $\mathbf{b}$, $\mathbf{f}$, $\mathbf{v}$ and $\mathbf{u}$) is defined as
\begin{equation}
\begin{aligned}
\mathbf{a}\left(\bm{\theta}\right)=\sqrt{\frac{1}{N_{x}N_{y}}}\left[1,\cdots,e^{j \pi\left(N_{x}-1\right) \sin \theta \cos \phi}\right]^T \otimes \\
\left[1, \cdots, e^{j \pi\left(N_{y}-1\right) \sin \phi}\right]^T,
\end{aligned}
\end{equation}
where $\theta$ and $\phi$ are the azimuth and elevation angles, and $N_{x}$ and $N_{y}$ are the number of antennas along each dimension of the UPA, such that $N_{\text{T}}^{\text{gNB}}=N_{x} N_{y}$ for the s-gNB's transmit array, with analogous definitions for other arrays. 

The transmit beamforming vector $\mathbf{f}_n$ at the $n$th slot is designed based on a one-step prediction of the prior estimated angles, $\hat{\bm{\theta}}_{n | n-1}$, and is given by
\begin{equation}
\mathbf{f}_{n}=\mathbf{a}(\hat{\bm{\theta}}_{n | n-1}),
\end{equation}
where the prediction mechanism is detailed in Section. \ref{sensHO}. 

The initial angle estimate $\hat{\theta}_{0}$ is obtained using the inherent sensing capability of the gNB, which detects incoming vehicles and forms an initial beam toward the detected direction for synchronization and access. This initialization procedure follows our previous work~\cite{li2024frame}, where the sensing-based target detection and initial access were detailed. In this work, we concentrate on the handover process within a scenario featuring a single moving vehicle in the NR-V2I network operating in connected mode. By filtering out zero relative velocities in the Doppler domain, the moving vehicle can be isolated from stationary scatterers. Following the signal processing techniques in~\cite{li2024frame}, the critical motion-related parameters, specifically, the range $d_n$, radial velocity $v_n$, azimuth angle $\theta_n$, and elevation angle $\phi_n$, may be readily estimated from the received echo signals.

\section{Classic Handover in NR}\label{classicHO}

Efficient handover mechanisms are paramount in NR FR2 deployments due to the smaller cell sizes inherent to higher frequencies, leading to an increased handover frequency compared to FR1 and LTE networks. This section details the frame structure and protocol considerations for intra-frequency, inter-cell classic handover in FR2 NR communication networks.

\subsection{Frame Structure}

NR employs Discontinuous Reception (DRX) in connected mode to optimize power usage at the UE. As defined in 3GPP TS 38.331 \cite{3gpp.38.331}, the DRX cycle comprises two phases: DRX-OnDuration, where the UE actively monitors the channel, and DRX-OffDuration, during which the UE enters a low-power state. The DRX-OnDuration involves monitoring downlink data from the s-gNB, control signals, and reference signals from both the s-gNB and neighboring candidate gNBs (c-gNBs). The durations of the DRX cycle and its active phase are configurable, allowing the network to tailor power savings to operational needs.

\begin{figure}[htbp]
\centering
\includegraphics[width=\columnwidth]{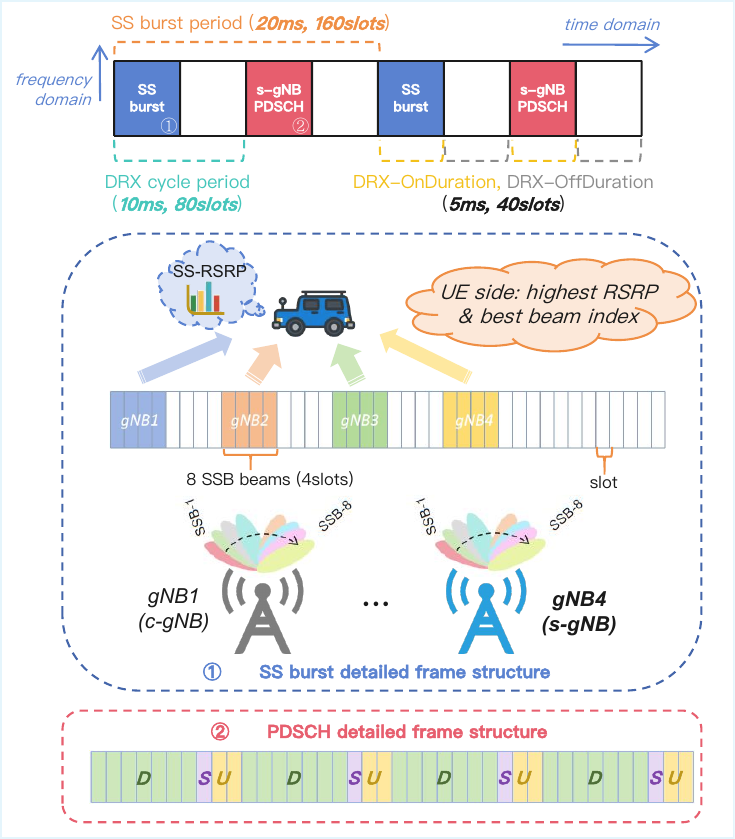}
\caption{Adopted frame structure.}
\label{comm_frame}
\end{figure}

For vehicular networks characterized by rapid movement, a shorter DRX cycle is advantageous. Here, we assume a DRX cycle of 10 ms (the length of one radio frame), with the DRX-OnDurationTimer set to 5 ms, allocating equal time to active monitoring and power conservation. During the active phase, the UE performs two critical functions: signal quality assessment and downlink data reception. 

Firstly, the UE monitors Synchronization Signal (SS) bursts to maintain synchronization with the s-gNB and evaluate neighboring cells. In NR, Synchronization Signal Blocks (SSBs), each containing the primary synchronization signal (PSS), secondary synchronization signal (SSS), and physical broadcast channel (PBCH), are grouped into SS bursts, which are transmitted periodically (e.g., every 20 ms). Each SSB corresponds to a distinct beam direction, enabling beam-based measurements and selection. These bursts from the serving and neighboring gNBs, time division multiplexed to avoid interference, are crucial in making potential handover decisions by identifying cells with stronger signals. 

Additionally, the UE receives downlink transmissions from the s-gNB via the Physical Downlink Shared Channel (PDSCH), ensuring uninterrupted communication. In this study, we adopt a commonly used 5G NR frame configuration, denoted as "DDDDDDDSUU", where each "D" indicates a downlink slot, "U" denotes an uplink slot, and "S" represents a special slot used for transitions between uplink and downlink transmissions, including a guard period~\cite{ATIS.3GPP.37.910.V1610}. To enhance reliability under typical high-mobility conditions in V2I scenarios, the Demodulation Reference Signal (DMRS) associated with the PDSCH employs the mapping type "A" with one additional DMRS symbol. Other reference signals such as CSI-RS and PTRS are not considered in this study. The adopted frame structure is illustrated in Fig.~\ref{comm_frame}.

\subsection{Triggering Event and Handover Command}

As a vehicle moves within the coverage area of the s-gNB, fluctuations in signal quality due to factors such as distance or physical obstructions can necessitate a handover to a neighboring gNB to maintain optimal service. As specified in TS 38.331~\cite{3gpp.38.331}, several events are defined as triggering events to help identify whether a handover is needed. Among these, Event A3 is commonly used. The triggering condition for Event A3 is met when the quality of reference signals from a neighboring gNB exceeds that of the s-gNB by a predefined offset. Reference Signal Received Power (RSRP) is a commonly used metric to quantify the quality of reference signals. The gNB that satisfies the triggering condition is designated as the target gNB (t-gNB).

\begin{figure*}[htbp]
\centering
\includegraphics[width=18cm]{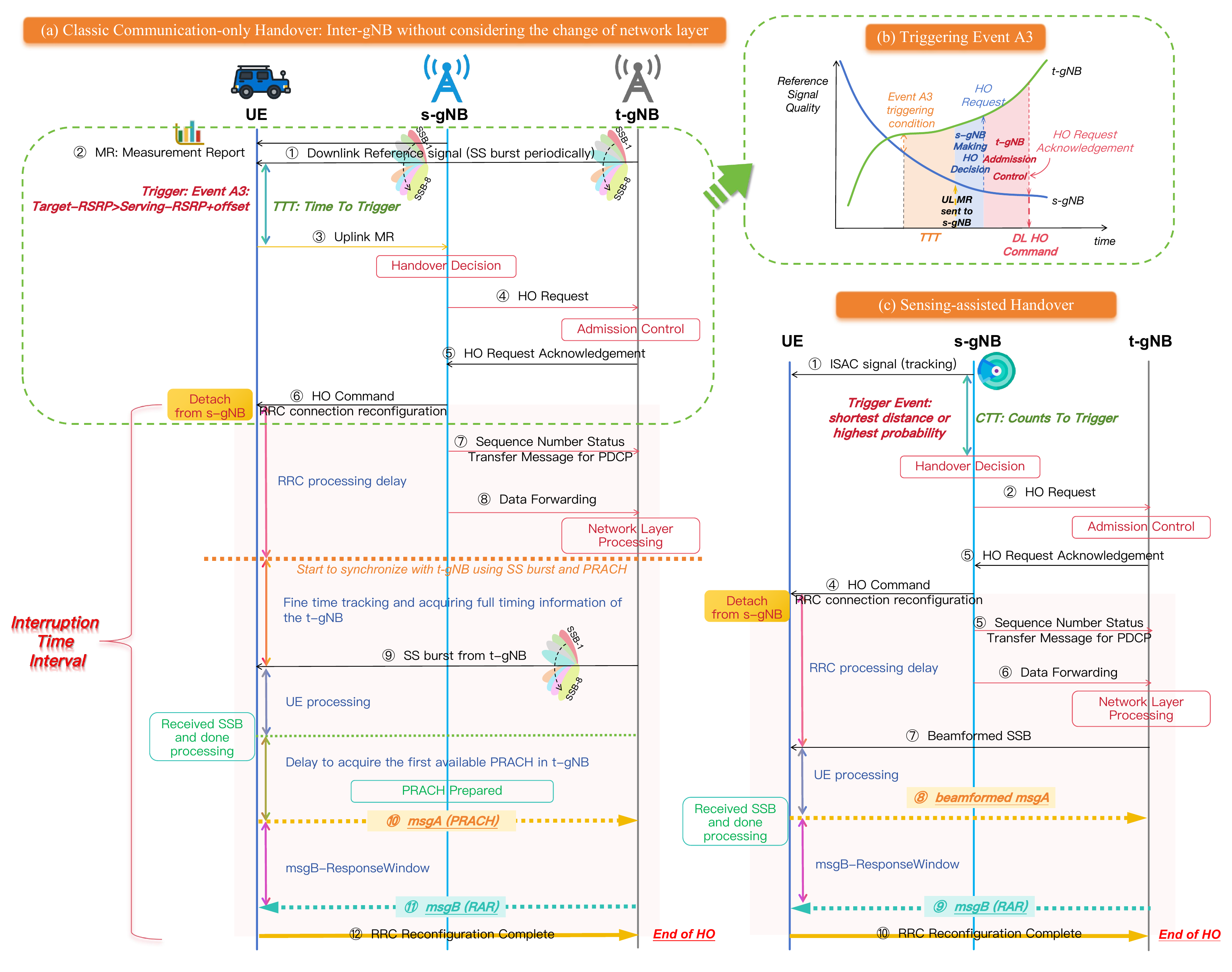}
\caption{Event-A3 triggered handover.}
\label{protocol}
\end{figure*}

The overall protocol flow for classic communication-only inter-gNB handover is presented in Fig.~\ref{protocol} (a) and (b). To avoid unnecessary handovers (e.g., ping-pong effects), the UE delays reporting until the Event A3 condition persists for a duration termed \textit{Time To Trigger (TTT)}. A well-chosen TTT balances stability and responsiveness: an overly long TTT risks handover failure, while a brief TTT may cause oscillation between gNBs. Once TTT elapses, the UE transmits a Measurement Report (MR) with RSRP values to the s-gNB via uplink. The s-gNB then assesses the MR, issues a handover request to the t-gNB, and awaits confirmation based on the resource availability of the t-gNB. Upon approval, the UE receives a handover command, which marks the start of the \textit{Interruption Interval}, a period without data exchange, and the transition to the t-gNB. The interruption interval, though ideally brief, encompasses essential preparatory steps for a successful handover. These steps, detailed in Fig.~\ref{protocol} (a), vary with different configurations listed in Table.~\ref{IT}, but are critical for synchronization and connectivity.

The interruption interval in an NR handover begins with a 10 ms Radio Resource Control (RRC) processing delay, as specified in 3GPP TS 38.331~\cite{3gpp.38.331}, during which the UE decodes and processes the RRC Connection Reconfiguration message. This step reconfigures Layer 2 and Layer 3 protocols, such as Medium Access Control (MAC), Packet Data Convergence Protocol (PDCP), and Radio Link Control (RLC), and activates security mechanisms like integrity protection and ciphering for the new link. Concurrently, the s-gNB transmits the Sequence Number (SN) Status Transfer Message and forwards buffered data packets to the t-gNB over the Xn interface, a backhaul process that does not extend the interruption time since it avoids air interface interaction. Next, the UE synchronizes with the t-gNB by detecting its SS burst, a critical process for frame timing alignment and limited to the SSB-based Measurement Timing Configuration (SMTC) period~\cite{3gpp.38.133}, which ranges from 5 to 160 ms depending on network configuration. Following synchronization, the UE completes internal processing, including RF retuning to the t-gNB's frequency, baseband adjustments, and SSB post-processing, within a maximum of 20 ms, as per TS 38.133~\cite{3gpp.38.133}. To reduce latency, a 2-step Contention-free Random Access (CFRA) procedure is utilized. The delay to the first Physical Random Access Channel (PRACH) opportunity, denoted $T_{\text{IU}}$, depends on the PRACH configuration index, periodicity (e.g., 10 to 40 ms), and the optimal SSB beam, which is indicated by the UE selecting a specific PRACH slot. Within a configured response window, the UE receives the Random Access Response (RAR, or msg-B), delivering the uplink resource grant and timing advance command to align uplink transmissions. The handover concludes when the UE sends the RRC Reconfiguration Complete message to the t-gNB over the allocated uplink resource, terminating the interruption interval and establishing the t-gNB as the new s-gNB.

\begin{table}[!ht]
    \caption{Components of interruption time} \label{IT} 
    \normalsize
    \centering
    \begin{tabular}{p{5.5cm} p{2.5cm}} 
    \hline
    \hline
    Components & Time \\
    \hline
    RRC processing delay & 10 ms\\
    Fine time tracking and acquiring full timing information of the t-gNB $T_{\Delta}$ & SMTC (max)\\
    UE processing $T_{\text{processing}}$ & 20 ms (max)\\
    Interruption uncertainty $T_{\text{IU}}$ & 170 ms (max) \\
    msgA (PRACH) & 1 or 2 slots\\
    msgB-ResponseWindow & 320 slots (max)\\
    msgB (RAR) & 1 or 2 slots\\
    RRC Reconfiguration Complete & 8 slots\\
    \hline
    Typical interruption time & 40--60 ms \\
    Worst-case interruption time & $\geq$100 ms \\
    \hline
    \hline
    \end{tabular}
\end{table}

In summary, the classic NR handover procedure, although standardized and robust, introduces considerable latency due to its multi-step architecture and dependency on periodic opportunities for synchronization and random access. The delay components accumulate sequentially and are sensitive to the network's configuration parameters. Even under favorable conditions with efficient PRACH timing and minimal UE processing, the total interruption duration typically ranges from 40 to 60 ms. In worst-case scenarios, especially under sparse PRACH configurations or unfavorable SMTC alignment, the total handover time can approach or surpass 100 ms. These latencies, coupled with high handover rate in FR2 V2I scenarios, highlight the pressing need for more predictive and low-latency handover frameworks, as addressed in the next section.

\section{Sensing-assisted Handover} \label{sensHO}

In this section, we introduce a novel sensing-assisted frame structure and protocol designed to minimize overhead and interruption time during handover in NR-V2I networks. We propose two sensing-assisted handover triggering schemes: \textit{Distance-based HO} and \textit{Probability-based HO}, which leverage kinematic parameters measured and estimated from the sensing capabilities inherent to gNBs, enhanced via Kalman filter (KF) and IMM-EKF tracking algorithms. Unlike traditional handover protocols that depend solely on the short and periodic SS resources of the frame and communication metrics, our approach integrates real-time sensing data to reduce latency and overhead, particularly in high-mobility V2I scenarios.

Conventional state evolution models, such as those for straight roads proposed in \cite{liu2020radar}, falter at intersections where vehicle trajectories (e.g., straight, left turn, or right turn) are uncertain. Furthermore, without prior knowledge of road geometry, prediction methods like those in \cite{meng2022vehicular} become impractical. Our schemes address these limitations by leveraging real-time kinematic data, enabling robust handover decisions without requiring road layout or vehicle intent information.

\begin{figure}[htbp]
\centering
\includegraphics[width=\columnwidth]{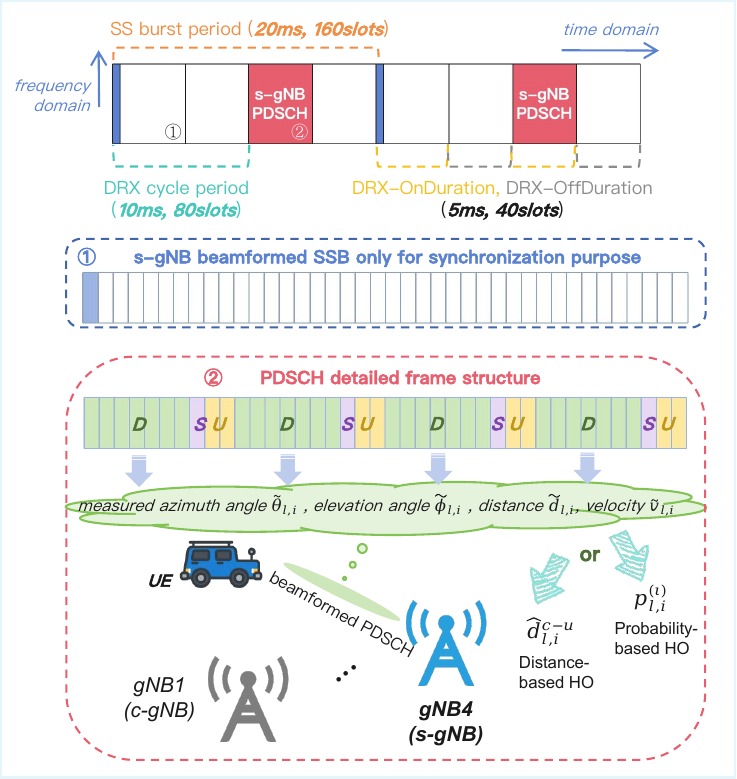}
\caption{Sensing-assisted frame structure.}
\label{ISACframe}
\end{figure}

\subsection{Frame Structure} 

In traditional NR systems, the UE monitors SSBs from neighboring gNBs during each SS burst period, assessing handover conditions based on RSRP. This process involves periodic beam sweeping, which incurs significant power consumption overhead, and relies on uplink feedback after a TTT interval, introducing delays. In contrast, our sensing-assisted handover protocol employs kinematic estimates derived from downlink signals, enabling the s-gNB to autonomously determine handovers without uplink feedback, thus reducing latency and overhead. The proposed frame structure for sensing-assisted handover is illustrated in Fig.~\ref{ISACframe}. 

As shown in Fig.~\ref{ISACframe}, each DRX-OnDuration enables the s-gNB to perform four radar measurements of the UE, producing estimates of azimuth angle $\hat{\theta}_{l,i}$, elevation angle $\hat{\phi}_{l,i}$, range $\hat{d}_{l,i}$, and radial velocity $\hat{v}_{l,i}$, where $i\in \{1,2,3,4\}$ denotes the measurement index within the $l$th PDSCH block. 
Unlike conventional handover, which yields a single RSRP comparison per SS burst period, our approach leveraging ISAC generates multiple kinematic data points per SS burst period. By employing tracking methods detailed in Sections \ref{blur} and \ref{IMM}, handover decisions will hinge on these parameters or directional probabilities rather than RSRP, relegating SSBs from beam sweeping purposes to synchronization purpose only.

\subsection{Distance-based Handover}\label{blur}

In the absence of road geometry or vehicle intent data, we propose a generalized state evolution model to track the UE's kinematic parameters
\begin{equation}
\begin{cases}
\theta_l = \theta_{l-1} + w_\theta, \\
\phi_l = \phi_{l-1} + w_\phi, \\
d_l = d_{l-1} - v_{l-1} \Delta T + w_d, \\
v_l = v_{l-1} + w_v,
\end{cases}
\end{equation}
where $\Delta T$ is the time interval between predictions, which equal to the period of SS burst in this context,  assumed sufficiently small to ensure minimal parameter shifts, keeping the tracking beam aligned with the UE for precise measurements. The state evolution model and the corresponding measurement model can be compactly expressed as:
\begin{equation} \label{compact}
\left\{\begin{array}{l}
\text {State Evolution Model: } \boldsymbol{\Upsilon}_l=\mathbf{h}\left(\boldsymbol{\Upsilon}_{l-1}\right)+\boldsymbol{w}_l, \\
\text {Measurement Model: } \Tilde{\boldsymbol{\Upsilon}}_{l,i}=\boldsymbol{\Upsilon}_l+\boldsymbol{\varepsilon}_{l,i},
\end{array}\right.
\end{equation}
where $\boldsymbol{\Upsilon}=[\theta, \phi, d, v]^T$ denotes the state variable vector, $\Tilde{\boldsymbol{\Upsilon}}=[\Tilde{\theta}, \Tilde{\phi}, \Tilde{d}, \Tilde{v}]^T$ denotes the measurement variable vector, as measured by the s-gNB following the radar measurement method in \cite{li2024frame}. The terms $\boldsymbol{w}=[w_\theta, w_\phi, w_d, w_v]^T$ and $\boldsymbol{\varepsilon}=[\varepsilon_\theta,\varepsilon_\phi, \varepsilon_d, \varepsilon_v]^T$ are the zero-mean Gaussian noises caused by approximation and measurement, respectively. Given the linearity of these models, a standard KF procedure can be employed for parameter estimation in this context and the estimated kinematic parameters after KF can be denoted as $\Hat{\boldsymbol{\Upsilon}} =[\Hat{\theta}, \Hat{\phi}, \Hat{d}, \Hat{v}]^T$.

With the s-gNB located at $[x^{\text{s}}, y^{\text{s}}, z^{\text{s}}]$ in the Cartesian coordinate system, the estimated location of the UE $[\hat{x}^{\text{u}}_{l,i}, \hat{y}^{\text{u}}_{l,i}, \hat{z}^{\text{u}}_{l,i}]$  at the $l$th PDSCH block and the $i$th measurement can be expressed as
\begin{equation}\label{7}
\begin{cases}
\hat{x}^{\text{u}}_{l,i} = x^{\text{s}} + \hat{d}_{l,i} \cos{\hat{\phi}_{l,i}} \cos{\hat{\theta}_{l,i}}, \\
\hat{y}^{\text{u}}_{l,i} = y^{\text{s}} + \hat{d}_{l,i} \cos{\hat{\phi}_{l,i}} \sin{\hat{\theta}_{l,i}}, \\
\hat{z}^{\text{u}}_{l,i} = z^{\text{s}} + \hat{d}_{l,i} \sin{\hat{\phi}_{l,i}}.
\end{cases}
\end{equation}

For a c-gNB $[x^{\text{c}}, y^{\text{c}}, z^{\text{c}}]$ is known by the s-gNB, the estimated distance to the UE is
\begin{equation}\label{8}
\hat{d}^{\text{c-u}}_{l,i}=\sqrt{(x^{\text{c}}-\hat{x}^{\text{u}}_{l,i})^2+(y^{\text{c}}-\hat{y}^{\text{u}}_{l,i})^2+(z^{\text{c}}-\hat{z}^{\text{u}}_{l,i})^2}.
\end{equation}

We introduce a new handover triggering event based on distance comparison, introducing a new triggering counter called Distance-Triggering-Counter (DTC) and a distance offset $d_\text{offset}$. If $\hat{d}^{\text{c-u}}_{l,i} + d_\text{offset}<\hat{d}_{l,i}$, the c-gNB becomes a potential t-gNB, incrementing the DTC. Replacing TTT in conventional communication handover, we propose a new parameter \textit{Counts To Trigger (CTT)} as a threshold for consecutive DTC increments from the same t-gNB. Handover is triggered when the DTC exceeds CTT. The detailed steps of triggering distance-based handover can be summarized in Algorithm \ref{alg:distanceHO}.

\begin{algorithm}[htbp]
\caption{Distance-based Handover}
\label{alg:distanceHO}
\begin{algorithmic}[1]
\Require Candidate gNB positions $\{(x^c,y^c,z^c)\}$, CTT threshold
\State Initialize DTC$~\leftarrow0$ for each neighboring gNB
\For{each DRX-OnDuration indexed by $l$}
  \For{$i=1$ to $4$ measurements}
    \State Obtain $\Hat{\boldsymbol{\Upsilon}}_{l,i}$ via KF
    \State Compute UE position $[\hat{x}^{\text{u}}_{l,i}, \hat{y}^{\text{u}}_{l,i}, \hat{z}^{\text{u}}_{l,i}]$ via Eq.~\eqref{7}
    \For{candidate gNB}
      \State Compute distance $\hat d^{\text{c-u}}_{l,i}$ via Eq.~\eqref{8}
      \If{$\hat{d}^{\text{c-u}}_{l,i} + d_\text{offset}<\hat{d}_{l,i}$}
        \State DTC$~\gets$ DTC$~+ 1$
      \Else
        \State DTC$~\gets 0$
      \EndIf
      \If{DTC$~\ge$ CTT}
        \State Trigger handover to t-gNB; \textbf{break}
      \EndIf
    \EndFor
    \If{handover triggered} \textbf{exit} loops
    \EndIf
  \EndFor
\EndFor
\end{algorithmic}
\end{algorithm}

\subsection{Probability-based Handover}\label{IMM}

At intersections, vehicles may proceed straight, turn right, or turn left. When the UE is moving away from the s-gNB, as indicated by the measured radial velocity, we employ IMM-EKF to estimate the probabilities of these maneuvers, thereby guiding handover to the t-gNB that aligns with the vehicle's likely direction. 

To accurately distinguish between the three potential intentions at an intersection, we expand the state vector to include Cartesian components. Specifically, the measured range $\Tilde{d}_{l,i}$ is decomposed into the x-component distance $\Tilde{d}_{l,i}^{x}=\Tilde{d}_{l,i}\cos{\Tilde{\phi}_{l,i}}\cos{\Tilde{\theta}_{l,i}}$ and the y-component distance $\Tilde{d}_{l,i}^{y}=\Tilde{d}_{l,i}\cos{\Tilde{\phi}_{l,i}}\sin{\Tilde{\theta}_{l,i}}$ with respect to the s-gNB in Cartesian coordinates. Similarly, the measured radial velocity $\Tilde{v}_{l,i}$ is also decomposed into x-component $\Tilde{v}_{l,i}^{x}=-\Tilde{v}_{l,i}\cos{\Tilde{\phi}_{l,i}}\cos{\Tilde{\theta}_{l,i}}$ and y-component $\Tilde{v}_{l,i}^{y}=-\Tilde{v}_{l,i}\cos{\Tilde{\phi}_{l,i}}\sin{\Tilde{\theta}_{l,i}}$. Therefore, for the driving direction illustrated in Fig. \ref{scenario} (c), the state evolution model for a straight maneuver can be derived from geometric relations as
\begin{equation}
\begin{cases}
\theta_l = \theta_{l-1} + \frac{v_{l-1}^{x} \Delta T \sin(\theta_{l-1})}{\cos^2(\theta_{l-1}) (d_{l-1}^{x} \cos(\theta_{l-1}) + d_{l-1}^{y} \sin(\theta_{l-1}))} + w_\theta, \\
\phi_l = \phi_{l-1} + w_\phi, \\
d_l^{x} = d_{l-1}^{x} + \frac{v_{l-1}^{x} \Delta T}{\cos^2(\theta_{l-1})} + w_{d_x}, \\
d_l^{y} = d_{l-1}^{y} + w_{d_y}, \\
v_l^{x} = v_{l-1}^{x} + w_{v_x}, \\
v_l^{y} = v_{l-1}^{y} + w_{v_y}.
\end{cases}
\end{equation}
In contrast, for turning maneuvers (left or right) we have:
\begin{equation}
\begin{cases}
\theta_l = \pi + \arctan\left( \frac{d_{l-1}^{y} \pm v_{l-1}^{y} \Delta T}{d_{l-1}^{x} + v_{l-1}^{x} \Delta T} \right) + w_\theta, \\
\phi_l = \phi_{l-1} + w_\phi, \\
d_l^{x} = d_{l-1}^{x} + v_{l-1}^{x} \Delta T + w_{d_x}, \\
d_l^{y} = d_{l-1}^{y} \pm v_{l-1}^{y} \Delta T + w_{d_y} ,\\
v_l^{x} = v_{l-1}^{x} + w_{v_x}, \\
v_l^{y} = v_{l-1}^{y} + w_{v_y},
\end{cases}
\end{equation}
where $\pm$ distinguishes right (+) and left (-) turns.

Let $\boldsymbol{\Xi}=[\theta, \phi, d^x, d^y, v^x, v^y]^T$ be the IMM state vector, $\Tilde{\boldsymbol{\Xi}}$ and $\hat{\boldsymbol{\Xi}}$ be the measurement and estimated vector. The compact form of the prediction and measurement is similar to Eq.~\ref{compact}. For each state evolution model $\iota \in \{1,2,3\}$ (going straight, turning left, tuning right), the standard IMM-EKF procedure iterates as follows:

{\it {1) State Prediction:}} 
\begin{equation}
\hat{\boldsymbol{\Xi}}_{l \mid l-1}^{\left(\iota\right)}=\mathbf{g}^{\left(\iota\right)}\left(\hat{\boldsymbol{\Xi}}_{l-1}^{\left(\iota\right)}\right),
\end{equation}
where $\mathbf{g}^{\left(\iota\right)}$ is the state evolution function for maneuver $\iota$.

{\it {2) Linearization:}} 
\begin{equation}
\mathbf{G}_{l-1}^{\left(\iota\right)}=\left.\frac{\partial \mathbf{g}^{\left(\iota\right)}}{\partial \boldsymbol{\Xi}}\right|_{\boldsymbol{\Xi}=\hat{\boldsymbol{\Xi}}_{l-1}^{\left(\iota\right)}}, \mathbf{H}_{l}^{\left(\iota\right)}=\mathbf{I}.
\end{equation}

{\it {3) MSE Matrix Prediction:}} 
\begin{equation}
\mathbf{M}_{l \mid l-1}^{\left(\iota\right)}=\mathbf{G}_{l-1}^{\left(\iota\right)} \mathbf{M}_{l-1}^{\left(\iota\right)} \mathbf{G}_{l-1}^{\left(\iota\right)H}+\mathbf{Q}_{s}^{\left(\iota\right)},
\end{equation}
where $\mathbf{Q}_{s}^{\left(\iota\right)}$ is the covariance matrix of system noise $\boldsymbol{w}^{\left(\iota\right)}$ for maneuver $\iota$.

{\it {4) Residual Update:}} 
\begin{equation}
\Tilde{\boldsymbol{\beth}}_{l,i}^{\left(\iota\right)}=\Tilde{\boldsymbol{\Xi}}_{l,i}-\hat{\boldsymbol{\Xi}}_{l \mid l-1}^{\left(\iota\right)}.
\end{equation}

{\it {5) Residual Covariance Calculation:}} 
\begin{equation}
\mathbf{S}_{l}^{\left(\iota\right)}=\mathbf{Q}_{m}^{\left(\iota\right)}+\mathbf{H}_{l}^{\left(\iota\right)} \mathbf{M}_{l \mid l-1}^{\left(\iota\right)} \mathbf{H}_{l}^{\left(\iota\right)H},
\end{equation}
where $\mathbf{Q}_{m}^{\left(\iota\right)}$ is the covariance matrix of measurement noise $\boldsymbol{\varepsilon}^{\left(\iota\right)}$ for maneuver $\iota$.

{\it {6) Kalman Gain Calculation:}} 
\begin{equation}
\mathbf{K}_{l}^{\left(\iota\right)}=\mathbf{M}_{l \mid l-1}^{\left(\iota\right)} \mathbf{H}_{l}^{\left(\iota\right)H}\left(\mathbf{S}_{l}^{\left(\iota\right)}\right)^{-1}.
\end{equation}

{\it {7) State Estimation:}} 
\begin{equation}
\hat{\boldsymbol{\Xi}}_{l,i}^{\left(\iota\right)}=\hat{\boldsymbol{\Xi}}_{l \mid l-1}^{\left(\iota\right)}+\mathbf{K}_{l}^{\left(\iota\right)}\Tilde{\boldsymbol{\beth}}_{l,i}^{\left(\iota\right)}.
\end{equation}

{\it {8) MSE Matrix Estimation:}} 
\begin{equation}
\mathbf{M}_{l}^{\left(\iota\right)}=\left(\mathbf{I}-\mathbf{K}_{l}^{\left(\iota\right)} \mathbf{H}_{l}^{\left(\iota\right)}\right) \mathbf{M}_{l \mid l-1}^{\left(\iota\right)}.
\end{equation}

{\it {9) Likelihood Update:}}
\begin{equation}
L_{l,i}^{\left(\iota\right)}=\exp{\left(-\frac{1}{2}\Tilde{\boldsymbol{\beth}}_{l,i}^{\left(\iota\right)T}\left(\mathbf{S}_{l}^{\left(\iota\right)}\right)^{-1}\Tilde{\boldsymbol{\beth}}_{l,i}^{\left(\iota\right)}\right)} \left|2 \pi \mathbf{S}_{l}^{\left(\iota\right)} \right|^{-1/2}.
\end{equation}

{\it {10) Probability Update:}}
\begin{equation}
p_{l,i}^{\left(\iota\right)}=\frac{\left(\sum_{\kappa}{\varpi_{\kappa ,\iota}p_{l-1}^{\left(\kappa\right)}}\right)L_{l,i}^{\left(\iota\right)}}{\sum_{\iota}{\left(\sum_{\kappa}{\varpi_{\kappa ,\iota}p_{l-1}^{\left(\kappa\right)}}\right)L_{l,i}^{\left(\iota\right)}}},
\end{equation}
where $\varpi$ denotes the predefined transition probability matrix, $\varpi_{\kappa,\iota}$ is the transition probability from model $\kappa$ to $\iota$, $p_{l-1}^{\left(\kappa\right)}$ represents the averaged probability of maneuver $\kappa$ across estimations in the $l-1$th PDSCH block, respectively.

{\it {11) State Update:}}
\begin{equation}
\hat{\boldsymbol{\Xi}}_{l,i}=\sum_{\iota}{\hat{\boldsymbol{\Xi}}_{l,i}^{\left(\iota\right)}p_{l,i}^{\left(\iota\right)}},  ~\hat{\boldsymbol{\Xi}}_{l}^{\left(\iota\right)}=\overline{\hat{\boldsymbol{\Xi}}_{l,i}^{(\iota)}},
\end{equation}
where the overline denotes averaging over index $i$.

The s-gNB computes the probability of each maneuver using measurements from each PDSCH block. 
These probabilities enable the s-gNB to infer the vehicle's likely direction and select the appropriate t-gNB for handover. 
To support this process, we define a new triggering counter called Probability-Triggering-Counter (PTC). 
It is noted that, unlike threshold-based schemes, the proposed method does not rely on a fixed probability threshold value; instead, the decision stability is ensured through the Counts-To-Trigger (CTT) parameter.  Specifically:
\begin{itemize}
    \item \textbf{Turning Maneuver}: If the probability of a turning model (either left or right) exceeds that of the straight model, the PTC is incremented. When the PTC is incremented consecutively and exceeds CTT threshold, the s-gNB triggers handover to the t-gNB aligned with the turn.
    \item \textbf{Straight Motion}: If the straight model consistently exhibits the highest probability, handover is initiated when the estimated distance to the straight-path t-gNB becomes shorter than that to the s-gNB, confirmed after CTT measurements.
\end{itemize}

This approach mitigates redundant handovers by ensuring that directional intent is validated over multiple measurements, reducing unnecessary switches and associated interruption time. The detailed steps of triggering probability-based handover can be summarized in Algorithm~\ref{alg:probabilityHO}.

\begin{algorithm}[htbp]
\caption{Probability-based Handover}
\label{alg:probabilityHO}
\begin{algorithmic}[1]
\Require Transition probability matrix $\varpi$, maneuver models $\iota \in \{\text{straight}, \text{left}, \text{right}\}$, CTT threshold, candidate gNB positions $\{(x^c, y^c, z^c)\}$
\State Initialize model probabilities $p_0^{(\iota)}$ for each $\iota$
\State Initialize PTC$^{(\iota)} \gets 0$ for each $\iota$
\For{each DRX-OnDuration indexed by $l$}
  \For{$i=1$ to $4$ measurements}
    \State Obtain $\Hat{\boldsymbol{\Xi}}_{l,i}$ and $p_{l,i}^{(\iota)}$ via IMM-EKF
    \State Determine dominant model $\iota^* \gets \arg\max_{\iota} {p}_{l,i}^{(\iota)}$
  \If{$\iota^*$ is left or right}
    \State PTC$^{(\iota^*)} \gets$ PTC$^{(\iota^*)} + 1$
  \ElsIf{$\iota^*$ is straight}
    \State Compute distance $\hat{d}^{\text{c-u}}_{l}$ to straight-path c-gNB
    \If{$\hat{d}^{\text{c-u}}_{l,i} + d_\text{offset}<\hat{d}_{l,i}$}
      \State PTC$^{(\iota^*)} \gets$ PTC$^{(\iota^*)} + 1$
    \EndIf
  \EndIf
  \If{PTC$^{(\iota^*)} \ge$ CTT}
      \State Trigger HO to $\iota^*$-aligned t-gNB; \textbf{break} \& \textbf{exit}
    \EndIf
\EndFor
\EndFor
\end{algorithmic}
\end{algorithm}

\subsection{Interruption Time of Sensing-based Handover}\label{senspro}

While the sensing-based handover retains the RRC processing delay and UE processing time, it significantly reduces synchronization and access delays:
\begin{itemize}
    \item \textbf{DL SSB Transmission}: In conventional handovers, the UE waits for the next periodic SS burst from the t-gNB, typically every 5--20 ms. In contrast, the t-gNB, informed of the UE's location via the handover request from the s-gNB, transmits a beamformed SSB immediately for synchronization purpose.
    \item \textbf{UL PRACH Transmission}: Instead of waiting for the next scheduled PRACH occasion to indicate the synchronization outcome and optimal beam pair, the UE is able to send PRACH preamble immediately after receiving the SSB.
\end{itemize}

The RAR and RRC Reconfiguration Complete message incur the same delays as conventional methods. Thus, the sensing-based method reduces the total interruption time by approximately 10--30 ms compared to conventional handovers, enhancing connectivity in vehicular networks.

\section{Link-Level Simulations}\label{linkSim}

The simulation scenario illustrated in Fig.~\ref{scenario}(a) is constructed using real-world geographical data extracted from OpenStreetMap (OSM), centered around the Bao'an Stadium area in Shenzhen, China. A total of nine gNBs are deployed along the roadside at varying heights: some mounted on building rooftops, while others are placed at street level to mimic realistic urban deployment. Six distinct vehicular routes are defined in accordance with traffic rules, resulting in varying simulation durations across different trajectories. The vehicle is assumed to be already in connected mode with a s-gNB and is moving toward the edge of its coverage area. At each road intersection, the vehicle may take one of multiple possible directions and experience possible handovers.

The communication channel between the s-gNB and the vehicle evolves dynamically over time, comprising both LoS and NLoS paths. Beamforming is implemented at both the gNB side and the UE side. In the communication-only V2I baseline, beamforming is codebook-based using a predefined set of eight beams. In contrast, the sensing-assisted V2I configuration leverages predicted angle information for beamforming, which offers enhanced directionality and precision over codebook-based schemes. It should be noted that the triggering algorithm also influences the beamforming gain, as different handover timings or t-gNB selections result in varying beam alignments. 
Hence, the performance comparison reflects the combined effect of predictive beamforming and proactive triggering within the integrated sensing-assisted framework.
 To improve link robustness, LDPC and CRC mechanisms compliant with 3GPP protocols are employed to facilitate error correction during demodulation. The target code rate is set to 0.658 in accordance with the specification in~\cite{3gpp.38.214}. A complete list of simulation parameters is provided in Table~\ref{Sim}, where $N_{\text{RB}}$ denotes the number of resource block and $Q_{\text{M}}$ denotes the modulation order.

\begin{table}[!ht]
    \caption{Parameters of simulation} \label{Sim} 
    \normalsize
    \centering
    \begin{tabular}{p{1.7cm} p{1.7cm} p{1.7cm} p{1.7cm}} 
    \hline
    \hline
    Parameter & Value & Parameter & Value\\
    \hline
    $f_c$ & $35\operatorname{GHz}$ & $T_{\text{max}}$ & $40\operatorname{s}$\\
    $\Delta f$ & $120\operatorname{kHz}$ & $\Delta T$ & $20\operatorname{ms}$\\
    $N_{\text{T}}^{\text{gNB}},N_{\text{R}}^{\text{gNB}}$ & $16\;(4\times 4)$ & $M_{\text{T}}^{\text{UE}},M_{\text{R}}^{\text{UE}}$ & $4\;(2\times 2)$\\
    $N_{\text{RB}}$ & $208$ & $Q_{\text{M}}$ & $4$\\
    $d_\text{offset}$ & $3\operatorname{m}$ & $\text{RSRP}_\text{offset}$ & $1\operatorname{dB}, 3\operatorname{dB}$\\
    $\text{SMTC}$ & $20\operatorname{ms}$ & $T_\text{processing}$ & $5\operatorname{ms}$\\
    msgA & $1\operatorname{slot}$ & msgB & $1\operatorname{slot}$\\
   \multicolumn{3}{p{5.1cm}}{msgB-ResponseWindow} & \multicolumn{1}{p{1.7cm}}{\(4\operatorname{slots}\)} \\
   \multicolumn{3}{p{5.1cm}}{PRACH Configuration Index} & \multicolumn{1}{p{1.7cm}}{\(70 - 88\)} \\
    \hline
    \end{tabular}
\end{table}

\subsection{Efficiency Comparison of Different HO Algorithms}

This subsection evaluates the handover efficiency of various algorithms in both communication-only and sensing-assisted V2I scenarios. In our simulations, each of the six vehicle routes is chosen with equal probability.

\begin{figure}[htbp]
\centering
\includegraphics[width=\columnwidth]{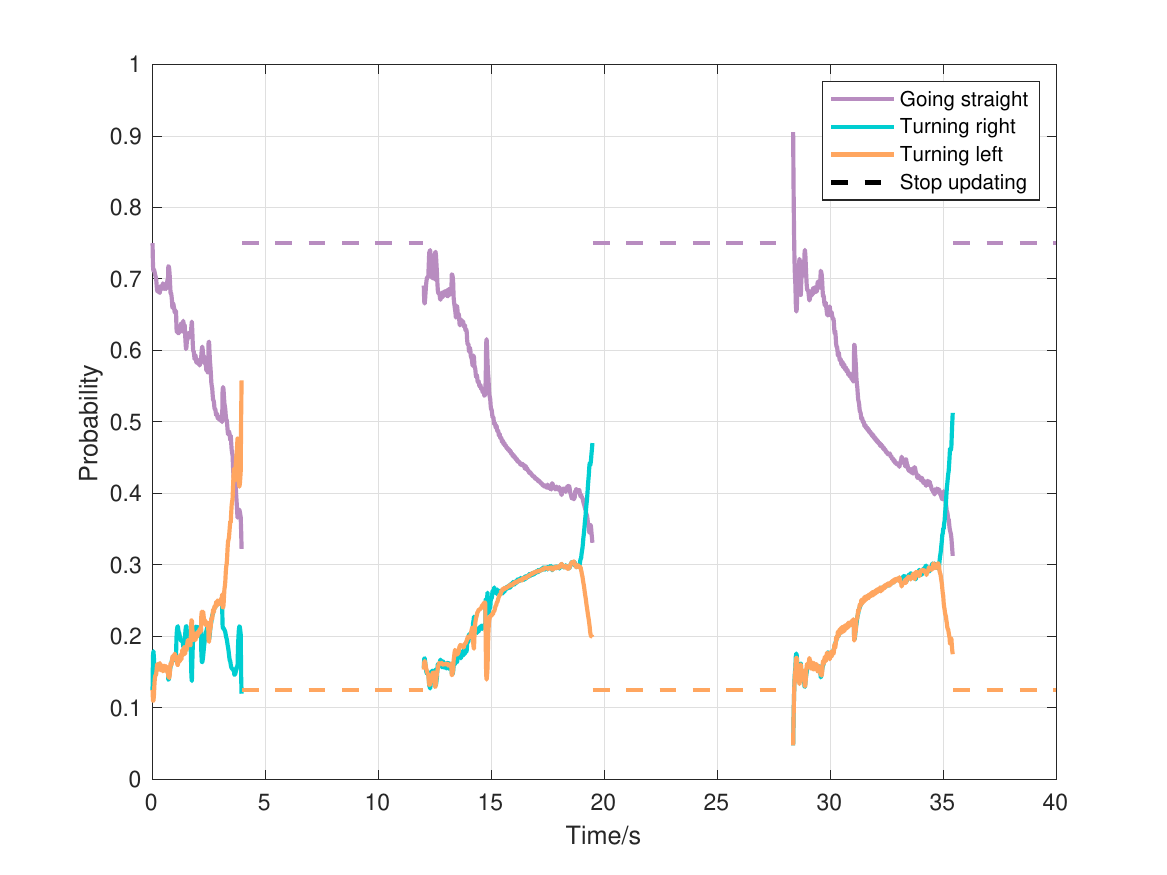}
\caption{Maneuver probabilities in probability-based handover (Route No.6).}
\label{IMM_prob}
\end{figure}

We first illustrate the evolution of maneuver probabilities along Route No.6 to demonstrate how the probability-based handover algorithm functions over time. In this setup, the CTT is configured as 16 counts, with the initial maneuver probability vector $p_0$ and the predefined transition probability matrix $\varpi$ given by 
\begin{equation}
p_0=\begin{bmatrix} \tfrac{3}{4} & \tfrac{1}{8} & \tfrac{1}{8}
\end{bmatrix},
\end{equation}
\begin{equation}
\varpi=\begin{bmatrix} 0.90 & 0.05 & 0.05\\ 0.05 & 0.90 & 0.05\\ 0.05 & 0.05 & 0.90
\end{bmatrix},
\end{equation}
where the dominant diagonal entries indicate the higher likelihood of a vehicle maintaining its current maneuver, while allowing for occasional transitions. The maneuver probabilities are updated only when the s-gNB detects that the vehicle is moving away from its coverage area, a condition inferred from the observed radial velocity. As shown in Fig.~\ref{IMM_prob}, the dashed lines represent time intervals where no update occurs (i.e., the vehicle is approaching the s-gNB). Once updates begin, the probability of going straight decreases while one alternative maneuver gains dominance, reflecting the design accuracy of the IMM-EKF state evolution model. When a maneuver’s probability exceeds the CTT threshold, a handover is triggered, and the UE connects to the t-gNB aligned with its predicted trajectory. For Route No.6, three intersections result in three successful handovers over the 40 s simulation period.

\begin{figure}[htbp]
\centering
\includegraphics[width=\columnwidth]{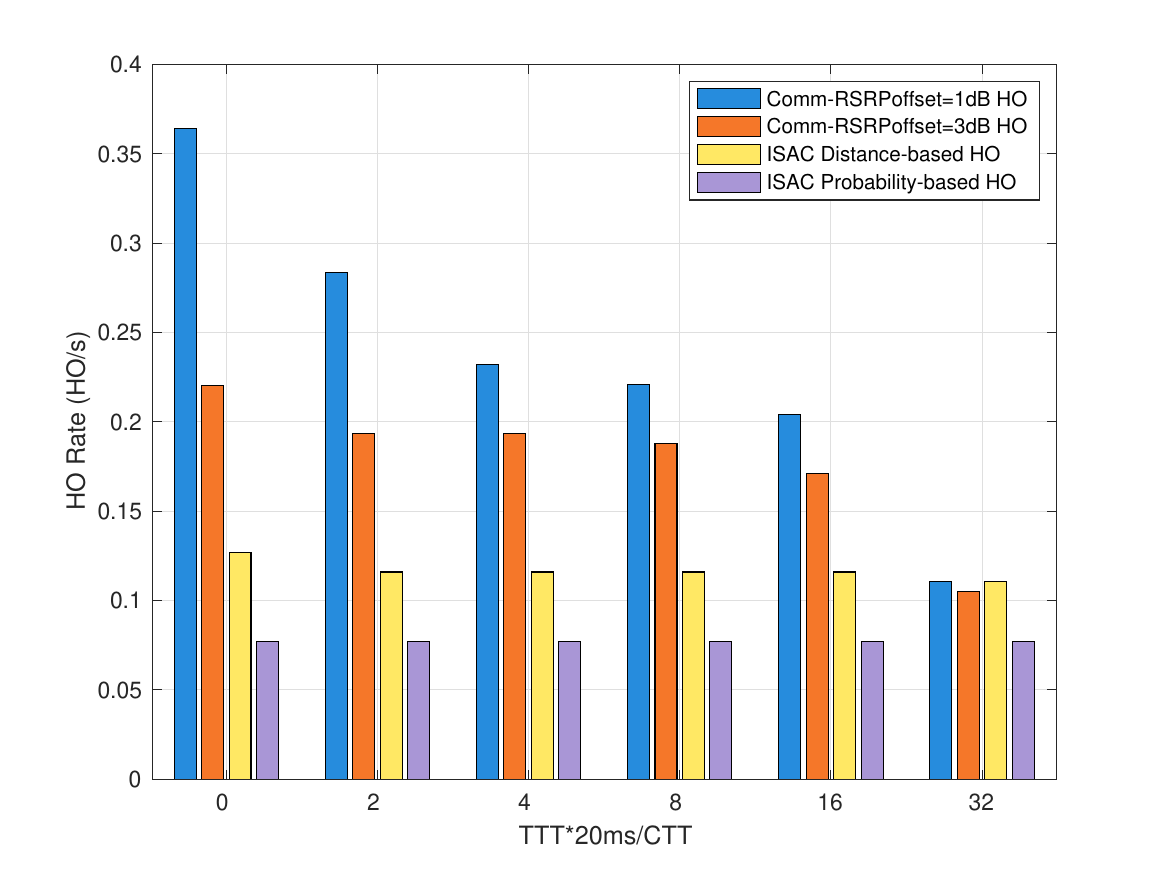}
\caption{HO rate across TTT/CTT values for different HO algorithms.}
\label{HO_rate}
\end{figure}

We define the handover rate as the number of handovers per unit time. As shown in Fig.~\ref{HO_rate}, communication-only schemes based on RSRP (Event A3) with smaller offsets (1 dB) and shorter TTT values result in higher handover rates. These frequent handovers are due to more easily satisfied triggering conditions. In contrast, sensing-based handover algorithms demonstrate more stable handover rates, largely because CTT is count-based rather than time-based, ensuring more consistent triggering durations and gNB selection behavior. Moreover, to align the sensing-assisted handover using CTT with the classic handover using TTT, it needs to be noticed that, for example, CTT=~4 accumulates 4 kinematic measurements over 20-40 ms, while 4 RSRP comparisons in the conventional scheme require a TTT of 4 SS burst periods, totaling 80 ms. This difference highlights the higher granularity of the sensing-assisted approach, ensuring a fair comparison over comparable decision windows.

\begin{figure}[htbp]
\centering
\includegraphics[width=\columnwidth]{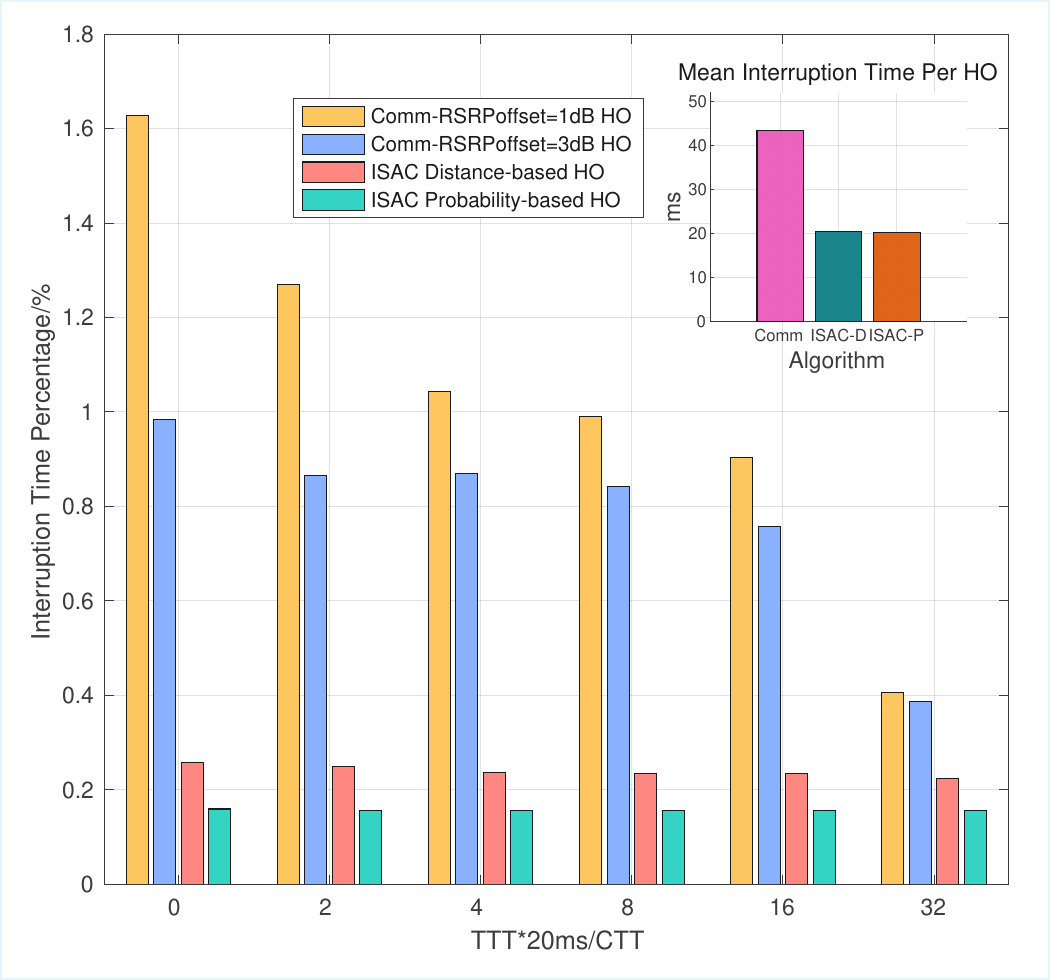}
\caption{Interruption time as a percentage of simulation time across TTT/CTT values for different HO algorithms.}
\label{HO_interruption}
\end{figure}

\begin{figure*}[htbp]
\centering
\includegraphics[width=18cm]{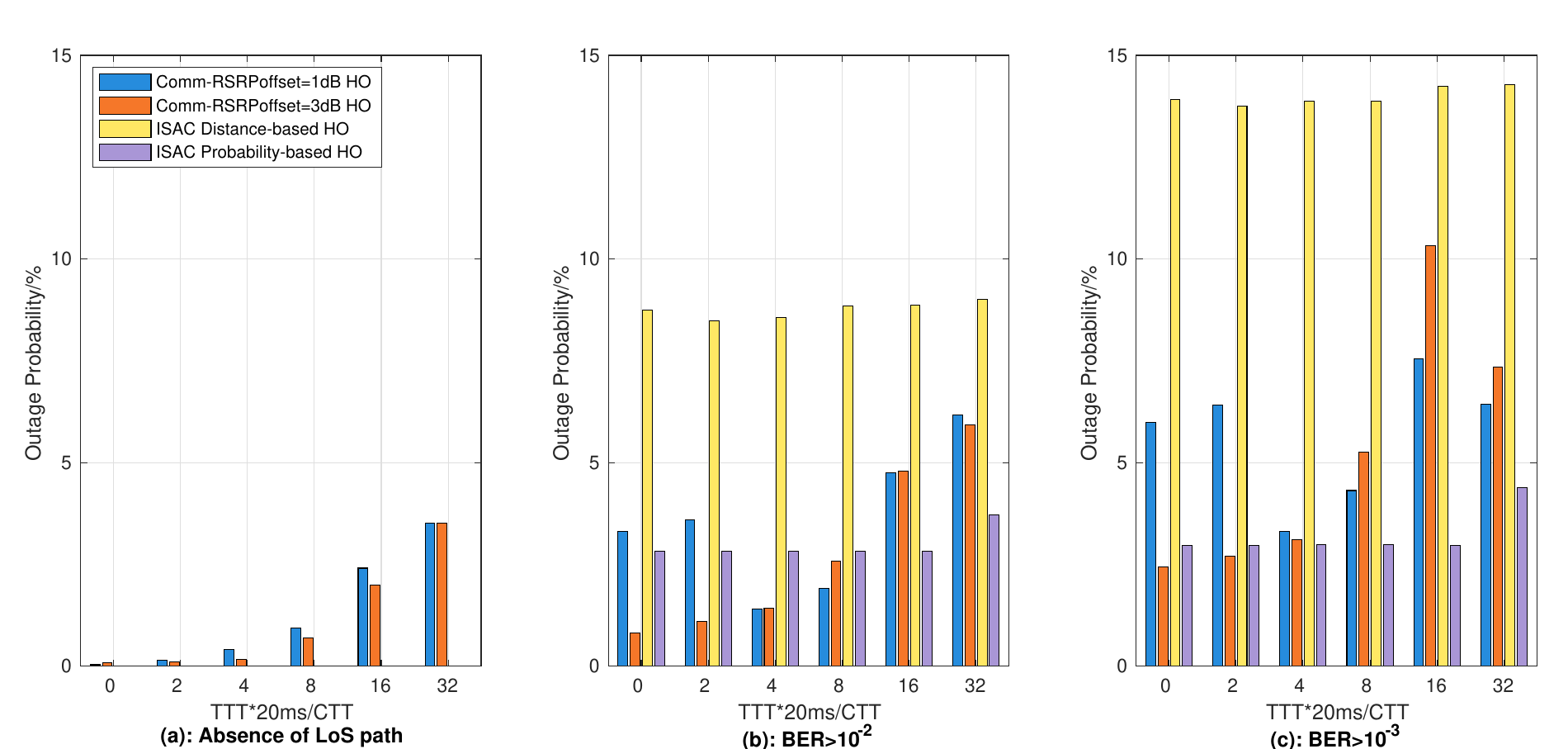}
\caption{Outage probability under different TTT/CTT of different algorithms for HO.}
\label{HO_Outage}
\end{figure*}

\begin{table*}[htbp]
\centering
\small % Slightly reduce font size for compactness
\setlength{\tabcolsep}{4pt} % Reduce column spacing to fit double-column width
\renewcommand{\arraystretch}{1.2} % Increase row spacing for readability
\caption{Bit error rate (BER) comparison for handover algorithms across TTT/CTT values.}
\label{tab:ber_comparison}
\begin{tabular}{l *{6}{S}}
\toprule
\textbf{HO Algorithm} & \multicolumn{6}{c}{\textbf{TTT($\mathbf{\times 20}$ms)/CTT}} \\
\cmidrule(lr){2-7}
& {0} & {2} & {4} & {8} & {16} & {32} \\
\midrule
 Comm-RSRPoffset=1dB & \num{3.23e-4} & \textbf{\num{3.31e-4}} & \textbf{\num{9.14e-5}} & \textbf{\num{2.21e-4}} & \num{1.50e-3} & \num{2.12e-3} \\
 Comm-RSRPoffset=3dB & \textbf{\num{2.17e-4}} & \num{3.95e-4} & \num{5.74e-4} & \num{9.91e-4} & \num{1.93e-3} & \num{2.11e-3} \\
 ISAC Distance-based & \num{4.30e-4} & \num{4.03e-4} & \num{4.18e-4} & \num{4.20e-4} & \num{5.47e-4} & \textbf{\num{4.32e-4}} \\
 ISAC Probability-based & \num{3.57e-4} & \num{3.45e-4} & \num{3.57e-4} & \num{3.61e-4} & \textbf{\num{3.51e-4}} & \num{7.62e-4} \\
\bottomrule
\end{tabular}
\end{table*}

One of the most critical performance metrics in high-mobility scenarios is the interruption time, during which data transmission is temporarily suspended as the UE transitions from the s-gNB to the t-gNB. According to the 3GPP standard and our prior analysis, this interval starts when the handover command is issued and ends when the RRC reconfiguration complete message is received by the t-gNB. During this window, no user data is exchanged, which degrades the user experience and reduces overall system throughput. To mitigate this, the proposed sensing-assisted protocol significantly shortens key components of the interruption interval. As illustrated in Figs.~\ref{protocol} and~\ref{ISACframe}, beamformed SSBs can be transmitted proactively by the t-gNB based on the predicted UE position. Likewise, the PRACH preamble can be sent immediately without waiting for the next PRACH occasion, thanks to angle-based alignment. These enhancements bypass conventional scheduling delays and minimize waiting time.

As illustrated in the inset of Fig.~\ref{HO_interruption}, the communication-only scheme results in an average interruption time of approximately 43 ms, while the sensing-assisted configuration reduces this to around 20 ms. This halving of the interruption interval translates into substantial improvements in maintaining continuous connectivity, especially to ensure consistent communication service and support latency-sensitive applications that demand minimal disruption. To assess the broader impact, we introduce the interruption time percentage in Fig.~\ref{HO_interruption}, defined as the ratio of total interruption time to the entire simulation duration. This metric captures the combined effects of both handover rate and the average time duration of each interruption. As shown in the figure, the interruption time percentage decreases with either a higher RSRP offset or the adoption of the sensing-assisted scheme. This improvement is attributed to both fewer handovers and shorter interruption durations per handover.

Fig.~\ref{HO_Outage} presents the outage probability under three definitions: absence of a LoS path, BER$>10^{-2}$, and BER$>10^{-3}$. Communication-only schemes experience a higher LoS-based outage due to their dependence on RSRP alone. Distance-based handover scheme results in slightly elevated BER-based outage, likely due to premature or mistimed handovers when only geometric distance is used. In contrast, the probability-based handover algorithm achieves balanced performance, resulting in consistently lower outage across all metrics.

\subsection{Communication Performance Comparison during Whole Process}

\begin{figure}[htbp]
\centering
\includegraphics[width=\columnwidth]{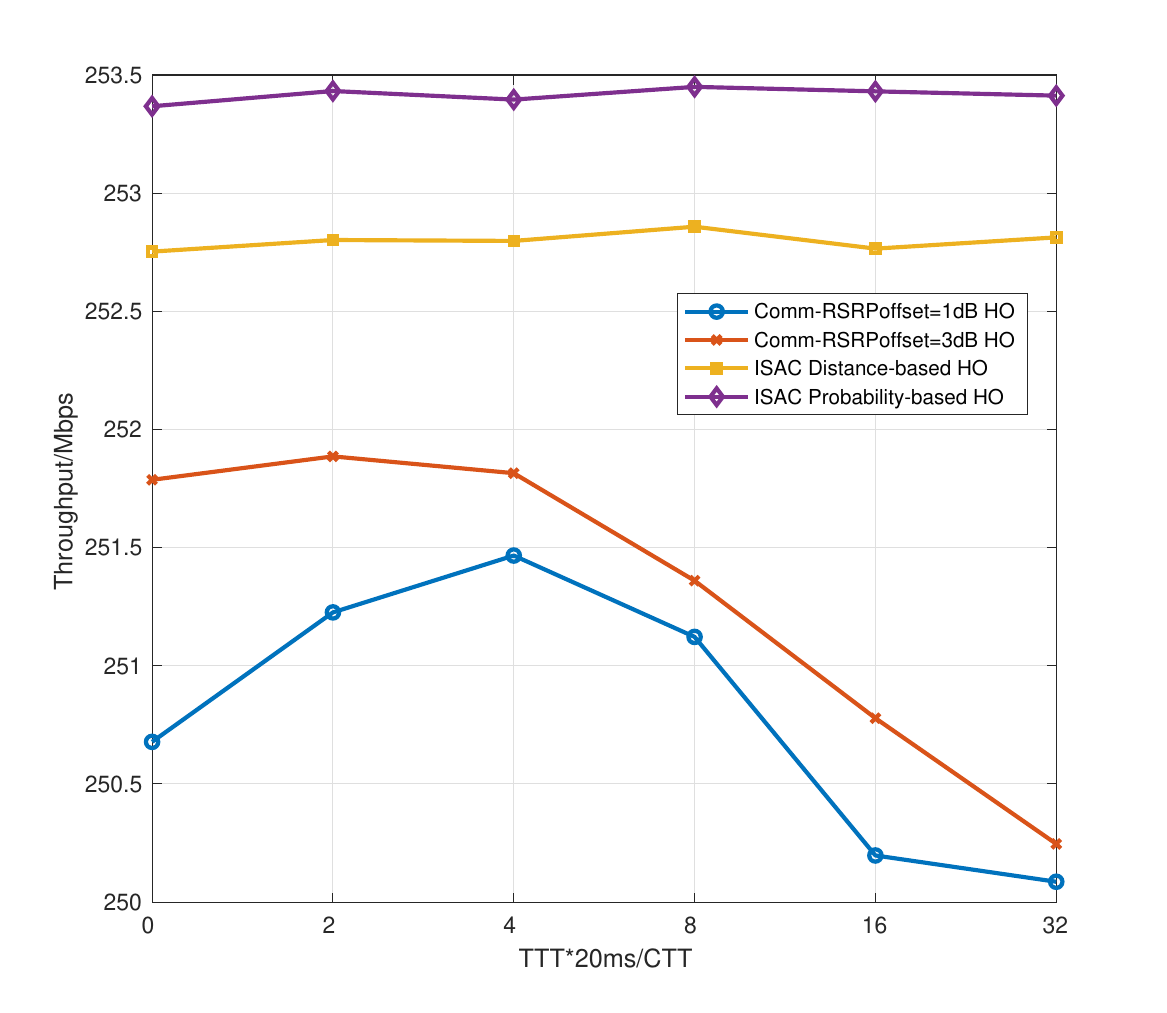}
\caption{Throughput comparison for handover algorithms across TTT/CTT values.}
\label{tp}
\end{figure}

Table~\ref{tab:ber_comparison} reports the average BER for each handover algorithm across TTT or CTT values. In general, communication-only handovers with a 1 dB offset perform best at low TTTs due to their aggressive strategy, which is quickly switching to stronger signals and minimizing exposure to degrading channels. However, this comes at the cost of frequent handovers, as seen in Fig.~\ref{HO_rate}, and consequently more interruptions. As TTT increases, communication-only systems may begin to miss timely handovers, especially when the UE approaches coverage boundaries, resulting in degraded BER. In contrast, sensing-assisted approaches show more robust performance as CTT increases. This can be attributed to the fact that CTT-based handover decisions are less susceptible to timing misalignments and rely on repeated spatial or probability confirmations of handover necessity. Additionally, sensing allows for more precise beamforming, further reducing BER degradation even when fewer handovers occur. Notably, although BER values across all algorithms remain within tolerable limits for V2I networks, the sensing-assisted probability-based scheme consistently balances sensitivity and stability, providing competitive BER while avoiding unnecessary interruptions.

Although the distance-based triggering scheme includes a distance threshold to prevent ping-pong effects, it may still exhibit slightly earlier triggering compared with the probability-based method as shown in Fig.~\ref{HO_Outage} and Table~\ref{tab:ber_comparison}. This behavior reflects its design philosophy, which prioritizes responsiveness by reacting promptly when the target gNB becomes closer than the serving one. The CTT parameter provides an adjustable tradeoff between responsiveness and robustness: increasing CTT delays triggering and improves stability, whereas smaller values favor immediate response at the cost of possible early handovers. Despite this aggressiveness, the distance-based triggering scheme offers a clear advantage in simplicity and implementability, requiring no uplink feedback or complicated probability calculations as in the IMM-EKF algorithm. This balance between fast reaction, robustness, and simplicity makes the scheme also feasible for real-time NR V2I deployment.

Fig.~\ref{tp} complements this observation by comparing system throughput across all settings. Throughput depends not only on BER but also on modulation order, redundancy overhead (e.g., reference signals), and interruption time during handovers. The sensing-enabled schemes, benefiting from reduced handover rate and shortened interruption intervals, deliver significantly better throughput under all CTT configurations. Interestingly, throughput in communication-only systems is non-monotonic with TTT. Shorter TTT improves BER due to more aggressive transitions but increases the number of handovers and hence the total interruption time, lowering effective data transfer. Conversely, longer TTT reduces handover rate but causes BER to worsen as the UE lingers in suboptimal channels. This tradeoff makes it challenging to tune TTT optimally. The CTT-based ISAC handover design avoids this issue by tying decisions to consistent motion patterns rather than fluctuating signal strength, allowing stable and high-throughput communication.

\section{Conclusion and Future Work}\label{Conclu}
In this paper, we have presented a comprehensive ISAC-enabled handover framework for 5G NR V2I networks, targeting the core challenges of frequent handovers, signaling overhead, and long interruption time in high-mobility urban environments. By exploiting the sensing capability of gNBs, we have proposed novel frame structures and two distinct distance-based and probability-based handover triggering algorithms, which rely on kinematic parameter estimation and maneuver prediction rather than conventional RSRP measurements.

Through detailed simulations in a realistic urban scenario, we have validated the advantages of the proposed framework. The sensing-assisted protocols have effectively reduced the average handover interruption time over 50\%, while decreasing handover rates and significantly improving system throughput. Overall, our approach have enabled more reliable and latency-sensitive V2I connectivity, especially critical for safety-critical and high-throughput vehicular applications.

Future work will focus on several directions to further enhance the proposed system. We plan to explore learning-based beam tracking and maneuver prediction models that can adapt to complex road layouts and dynamic traffic behaviors. Moreover, extending the framework to multi-vehicle settings and incorporating multi-user resource optimization remains a promising direction to fully realize the potential of ISAC in large-scale intelligent transportation systems. While the current handover protocol and frame structure are inherently scalable to multiple vehicles, further investigation is required to assess the effects of mutual sensing interference, coordinated resource allocation, and multi-target tracking accuracy under dense V2I deployments. Addressing these aspects will provide deeper insight into the framework’s scalability and robustness in realistic network conditions.

\bibliographystyle{IEEEtran}
\bibliography{IEEEabrv,ref}

\begin{IEEEbiography}[{\includegraphics[width=1in,height=1.25in,clip,keepaspectratio]{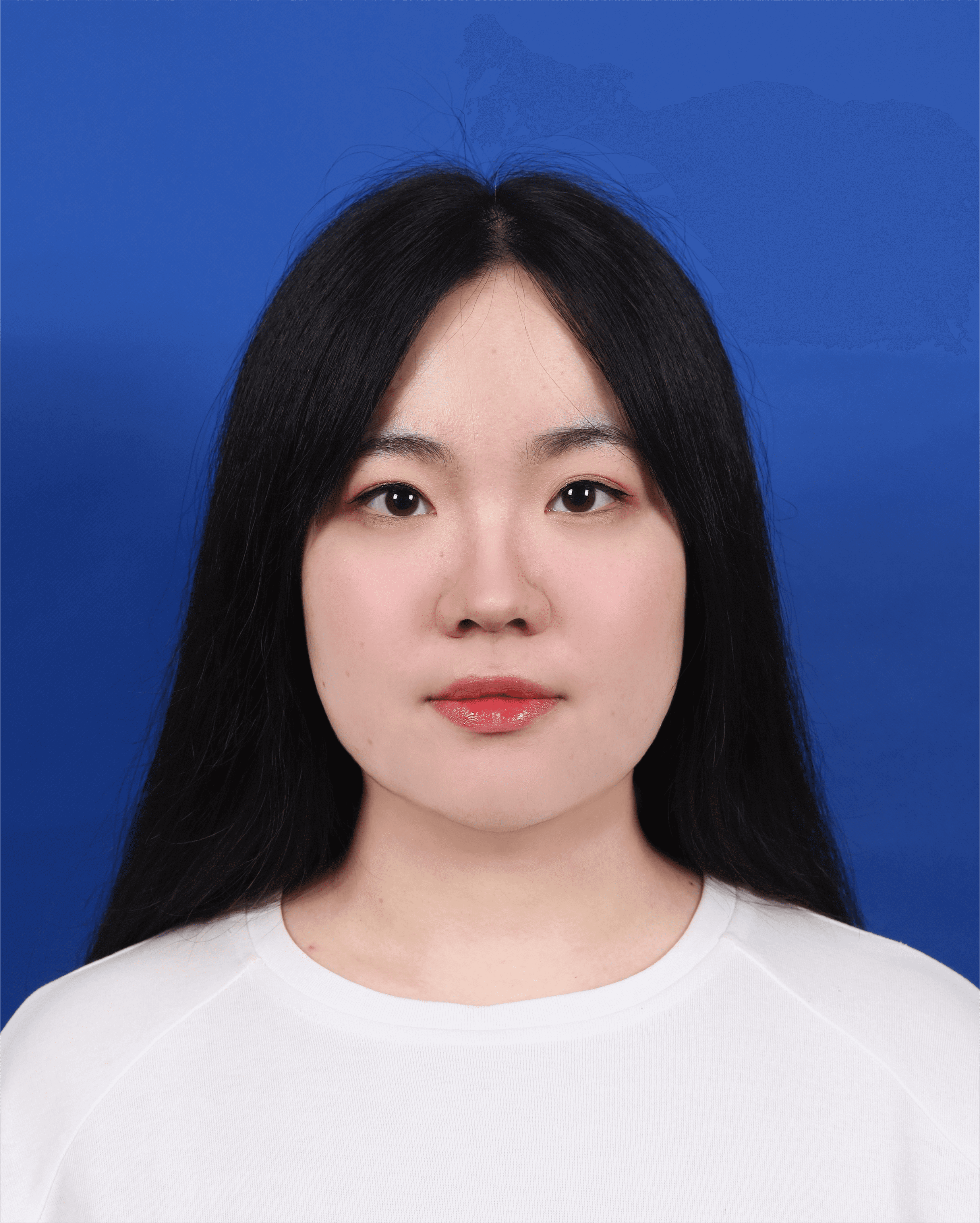}}]{Yunxin Li}(Graduate Student Member, IEEE) received the M.Eng. degree from the University of New South Wales, Australia. She is currently pursuing the Ph.D. degree with the School of Automation and Intelligent Manufacturing, Southern University of Science and Technology, Shenzhen, China. Since 2025, she has been a visiting Ph.D. researcher with the Networked Systems Group, Department of Electrical Engineering (ESAT), KU Leuven, Belgium. Her research interests include integrated sensing and communication (ISAC) and its application within the 5G NR protocol framework.
\end{IEEEbiography}

\begin{IEEEbiography}[{\includegraphics[width=1in,height=1.25in,clip,keepaspectratio]{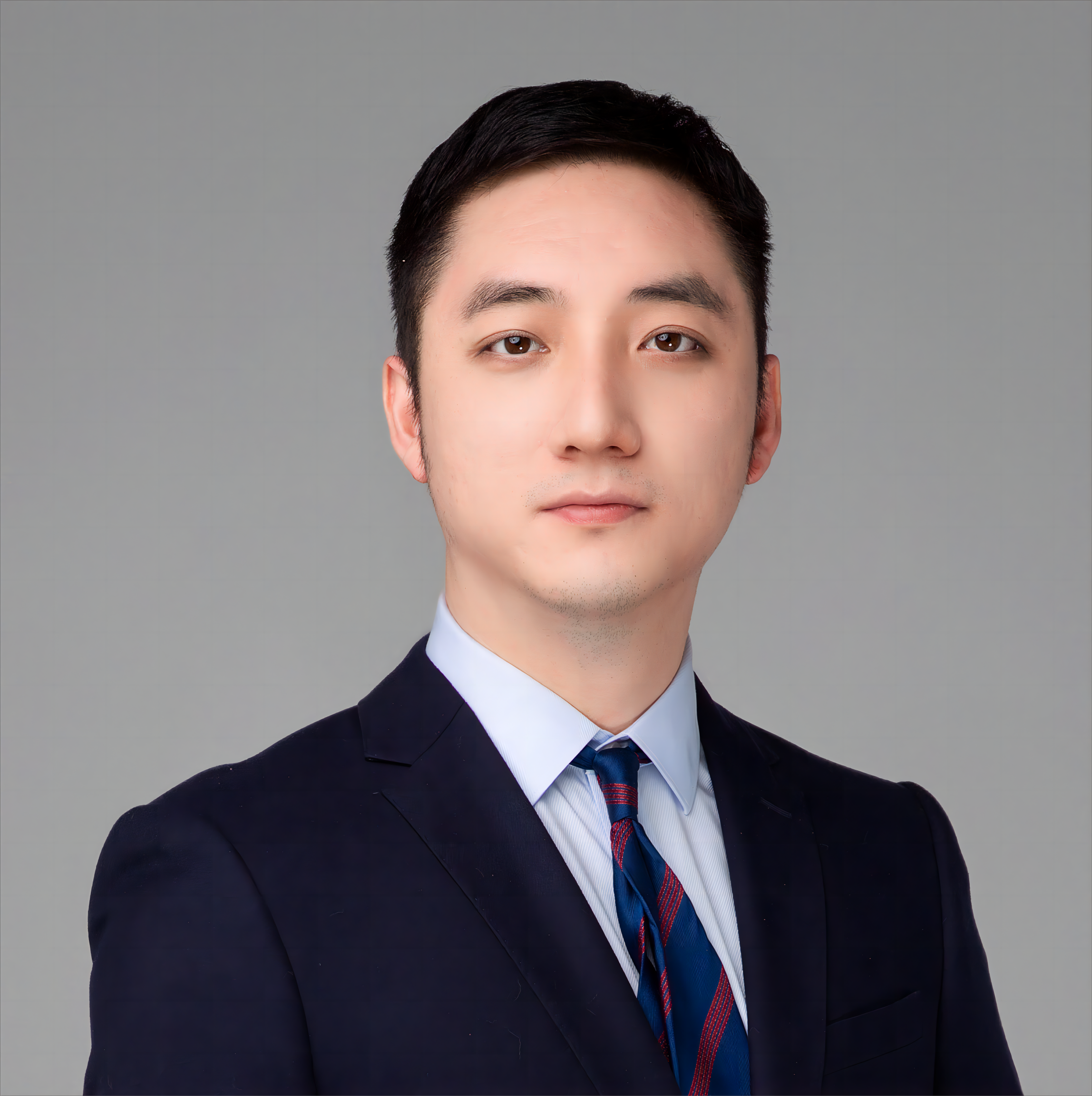}}]{Fan Liu}(Senior Member, IEEE) is currently a
Professor with the National Mobile Communications Research Laboratory, School of Information Science and Engineering, Southeast University, Nanjing, China. Prior to that, he was an Assistant Professor with the Southern University of Science
and Technology, Shenzhen, China, from 2020 to 2024. He received the Ph.D. and the BEng. degrees from Beijing Institute of Technology (BIT), Beijing, China, in 2018 and 2013, respectively. He has previously held academic positions in the University College London (UCL), London, UK, as a Visiting Researcher from 2016 to
2018, and a Marie Curie Research Fellow from 2018 to 2020.

Prof. Liu’s research interests lie in the general area of signal processing and wireless communications, and in particular in the area of Integrated Sensing and Communications (ISAC). He is the founding Academic Chair of the IEEE ComSoc ISAC Emerging Technology Initiative (ISAC-ETI), Vice Chair and founding member of the IEEE SPS ISAC Technical Working Group (ISAC-TWG), an elected member of the IEEE SPS Sensor Array and Multichannel Technical Committee (SAM-TC), an Associate Editor of the IEEE Transactions on Communications, IEEE Transactions on Mobile Computing, and IEEE Open Journal of Signal Processing, and a Guest Editor
of the IEEE Journal on Selected Areas in Communications, IEEE Wireless Communications, and IEEE Vehicular Technology Magazine. He was a TPC Co-Chair of the 2nd-4th IEEE Joint Communication and Sensing (JC\&S) Symposium, a Symposium Co-Chair for the IEEE ICC 2026 and IEEE GLOBECOM 2023, and a Track Co-Chair for the IEEE WCNC 2024. He is a member of the IMT-2030 (6G) ISAC Task Group. He was listed among the Clarivate Highly Cited Researcher in 2025, and among the Elsevier Highly-Cited Chinese Researchers from 2023 to 2024. He was a recipient of numerous Best Paper Awards, including the 2025 IEEE Communications Society \& Information Theory Society Joint Paper Award, 2024 IEEE Signal Processing Society Best Paper Award, 2024 IEEE Signal Processing Society Donald G. Fink Overview Paper Award, 2024 IEEE Communications Society Asia-Pacific Outstanding Paper Award, 2023 IEEE Communications Society Stephan O. Rice Prize, and 2021 IEEE Signal Processing Society Young
Author Best Paper Award.
\end{IEEEbiography}

\begin{IEEEbiography}[{\includegraphics[width=1in,height=1.25in,clip,keepaspectratio]{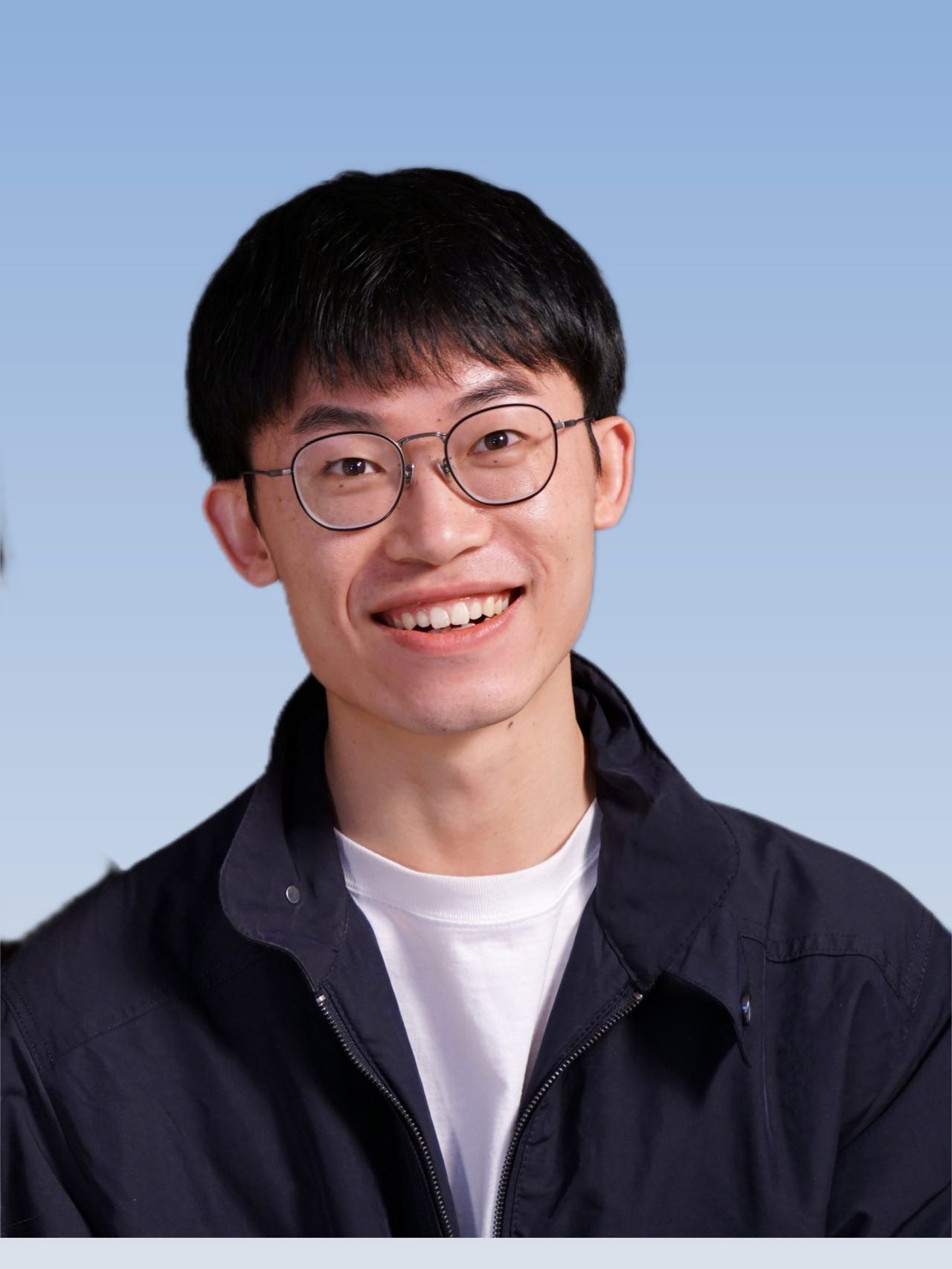}
}]{Haoqiu Xiong} (Graduate Student Member, IEEE) received the bachelor’s degree in optoelectronic information science and engineering from the Harbin Engineering University in 2019, and the M.Sc. by research in electronic engineering from the Southern University of Science and Technology in 2022. Since 2023, he has been a PhD student with the Networked Systems Group, Department of Electrical Engineering (ESAT), KU Leuven. His research interests include integrated sensing and communication, cell-free MIMO signal processing, and Wi-Gig sensing.
\end{IEEEbiography}

\begin{IEEEbiography}
[{\includegraphics[width=1in,height=1.25in,clip,keepaspectratio]{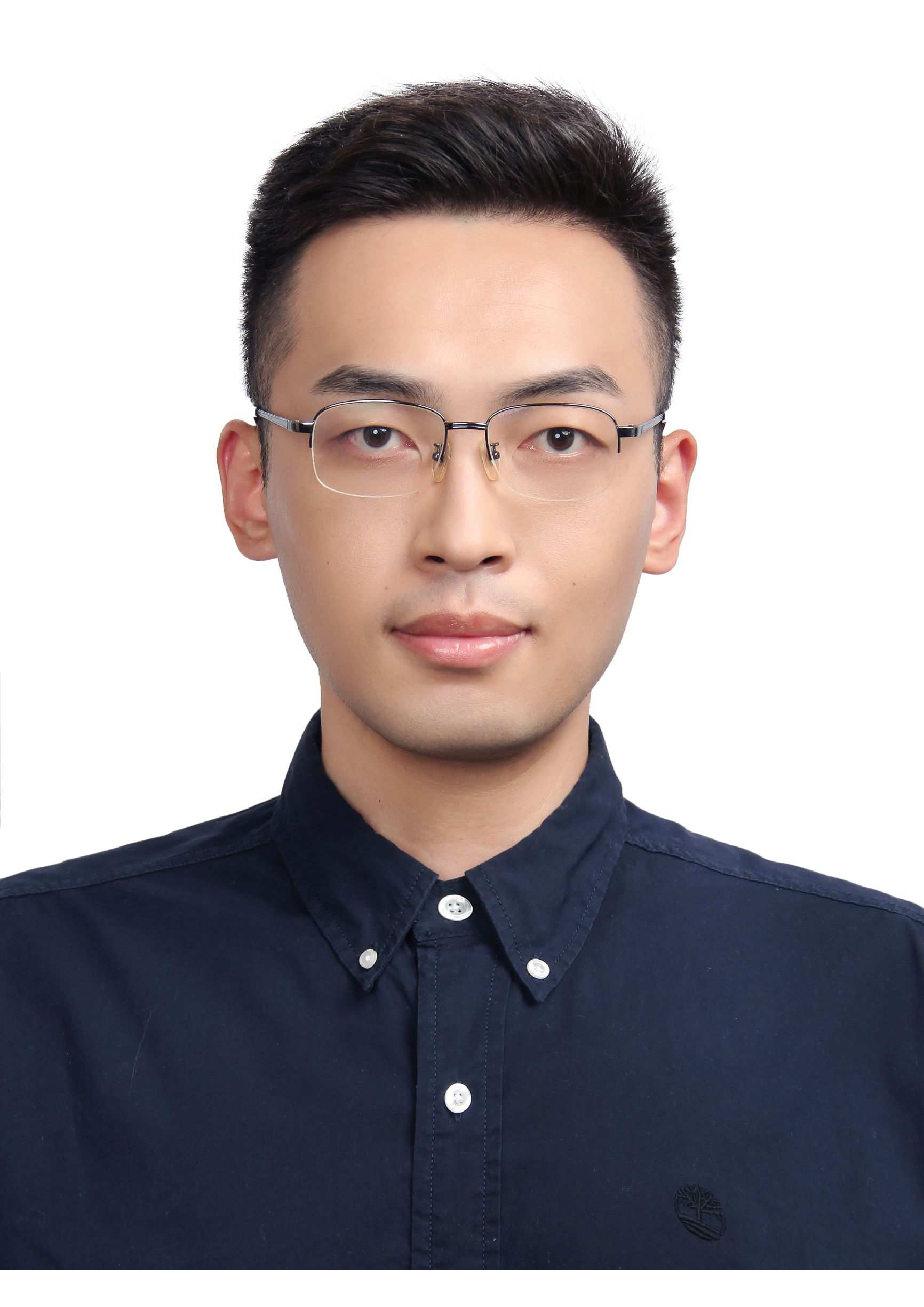}}]{Zhenkun Wang}(Senior Member, IEEE) received the Ph.D. degree in circuits and systems from Xidian University, Xi’an, China, in 2016. 

From 2017 to 2020, he was a Post-Doctoral Research Fellow with Nanyang Technological University, Singapore, and City University of Hong Kong, Hong Kong. He is currently an Assistant Professor with the School of Automation and Intelligent Manufacturing, Southern University of Science and Technology, Shenzhen, China. His research interests include AI-driven optimization, automated algorithm design, deep learning, and their applications.
\end{IEEEbiography}

\begin{IEEEbiography}[{\includegraphics[width=1in,height=1.25in,clip,keepaspectratio]{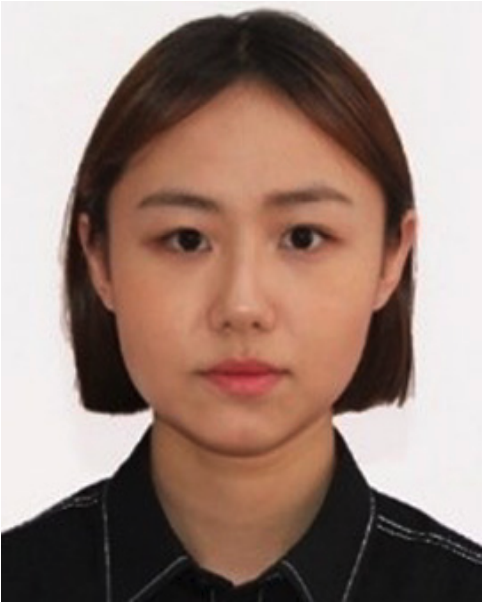}}]{Narengerile} (Member, IEEE) received the B.E.
degree in communication engineering from Sun Yat-sen University, Guangdong, China, in 2015, and the M.Sc. and Ph.D. degrees in digital communications from the University of Edinburgh, Edinburgh, U.K., in 2017 and 2022, respectively.
She is currently a Technical Engineer with Huawei Technologies Company Ltd. Her research interests include wireless communications, wireless local area network sensing, integrated sensing and communication, wireless communication standardization, and machine learning.
\end{IEEEbiography}

\begin{IEEEbiography}[{\includegraphics[width=1in,height=1.25in,clip,keepaspectratio]{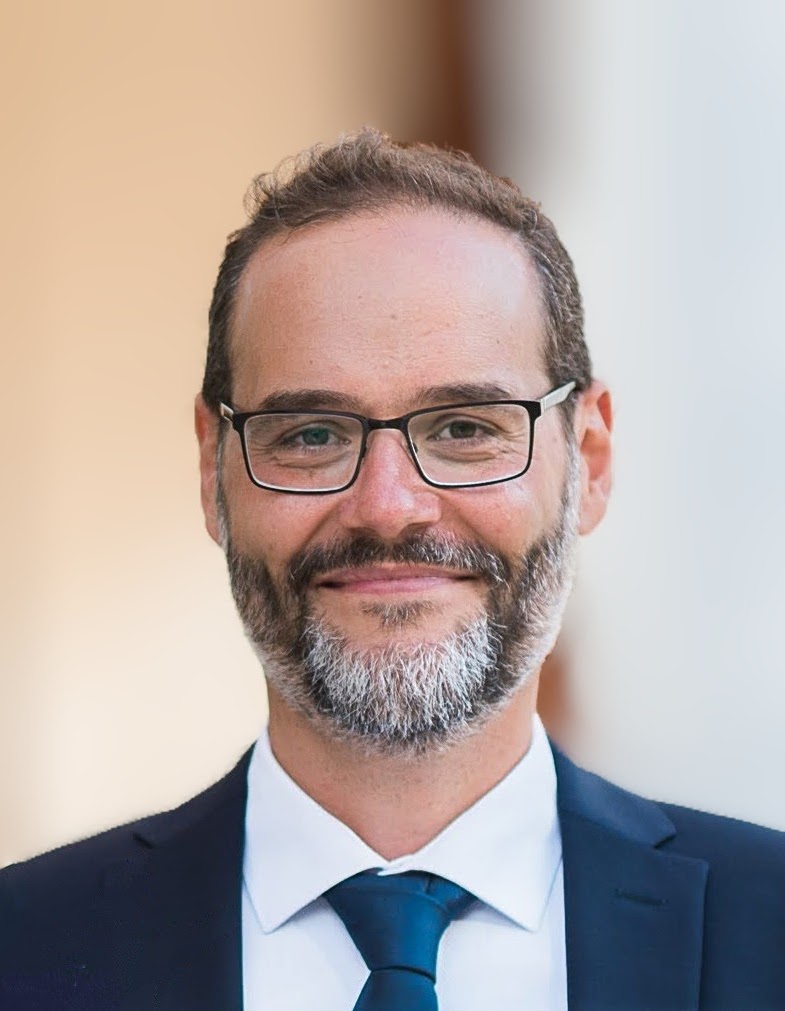}}]{Christos Masouros} (Fellow, IEEE) received the Diploma degree in Electrical and Computer Engineering from the University of Patras, Greece, in 2004, and MSc by research and PhD in Electrical and Electronic Engineering from the University of
Manchester, UK in 2006 and 2009 respectively. In 2008 he was a research intern at Philips Research Labs, UK, working on the LTE standards. Between 2009-2010 he was a Research Associate in the University of Manchester and between 2010-2012
a Research Fellow in Queen’s University Belfast. In 2012 he joined University College London as a Lecturer. He has held a Royal Academy of Engineering Research Fellowship between 2011-2016.

Since 2019 he is a Full Professor of Signal Processing and Wireless Communications in the Information and Communication Engineering research group, Dept. Electrical and Electronic Engineering, and affiliated with the Institute for Communications and Connected Systems, University College London. His research interests lie in the field of wireless communications
and signal processing with particular focus on Green Communications, Large Scale Antenna Systems, Integrated Sensing and Communications, interference mitigation techniques for MIMO and multicarrier communications. Between 2018-22 he was the Project Coordinator of the €4.2m EU H2020 ITN project PAINLESS, involving 12 EU partner universities and industries, towards
energy-autonomous networks. Between 2024-28 he will be the Scientific Coordinator of the €2.7m EU H2020 DN project ISLANDS, involving 19 EU partner universities and industries, towards next generation vehicular networks. He is a Fellow of the IEEE, Fellow of the Insitute of Electronic Engineers (IET), the Artificial Intelligence Industry Alliance (AIIA) and the Asia-Pacific Artificial Intelligence Association (AAIA). He was the recipient of the 2024 IEEE SPS Best Paper Award, the 2024 IEEE SPS Donald G. Fink Overview Paper Award, the 2023 IEEE ComSoc Stephen O. Rice Prize, co-recipient of the 2021 IEEE SPS Young Author Best Paper Award and the recipient of the Best Paper Awards in the IEEE GlobeCom 2015 and IEEE WCNC
2019 conferences. He is an IEEE ComSoc Distinguished lecturer 2024-2025, and his work on ISAC has been featured in the World Economic Forum’s report on the top 10 emerging technologies. He has been recognised as an Exemplary Editor for the IEEE Communications Letters, and as an Exemplary Reviewer for the IEEE Transactions on Communications. He is an Area Editor for IEEE Transactions on Wireless Communications, and Editor-atLarge for IEEE Open Journal of the Communications Society. He has been an Editor for IEEE Transactions on Communications, IEEE Transactions on Wireless Communications, the IEEE Open Journal of Signal Processing,
Associate Editor for IEEE Communications Letters, and a Guest Editor for a number of IEEE Journal on Selected Topics in Signal Processing issues. He is a founding member and Vice-Chair of the IEEE Emerging Technology Initiative on Integrated Sensing and Communications (SAC), Chair of the IEEE SPS ISAC Technical Working Group, and Chair of the IEEE Green
Communications \& Computing Technical Committee, Special Interest Group on Green ISAC. He is a member of the IEEE Standards Association Working Group on ISAC performance metrics, and a founding member of the ETSI ISG on ISAC. He is the TPC chair for the IEEE ICC 2024 Selected Areas in Communications (SAC) Track on ISAC, Chair of the IEEE PIMRC2024 Track 1 on PHY and Fundamentals, Chair of the ”Integrated Imaging and Communications” stream in IEEE CISA 2024, and TPC Co-Chair of IEEE VTC 2025.
\end{IEEEbiography}

\end{document}